\documentclass[preprint,3p,times]{elsarticle}

\usepackage{amssymb}
\usepackage{amsmath}
\usepackage[all,cmtip]{xy}
\usepackage{amsthm}
\usepackage{color}
\usepackage{enumitem}
\usepackage{tikz}
\usepackage{amsmath, mathrsfs}
\usepackage{appendix}
\usepackage[colorlinks=true]{hyperref}
\usepackage[hyphenbreaks]{breakurl}
\usepackage{diagbox} 
\usepackage[ruled,vlined]{algorithm2e}
\usepackage[normalem]{ulem}
\usepackage{booktabs}
\usepackage{pifont}
\usepackage{multirow}
\usepackage{xurl}

\input diagxy

\theoremstyle{plain}
  \newtheorem{thm}{Theorem}[section]
  \newtheorem{lem}[thm]{Lemma}
  \newtheorem{prop}[thm]{Proposition}
  
\theoremstyle{definition}
  \newtheorem{defn}[thm]{Definition}
  \newtheorem{exmp}[thm]{Example}
  \newtheorem{rem}[thm]{Remark}

\DeclareMathAlphabet{\mathcal}{OMS}{cmsy}{m}{n}

\makeatletter
\def\ps@pprintTitle{%
 \let\@oddhead\@empty
 \let\@evenhead\@empty
 \def\@oddfoot{\centerline{\thepage}}%
 \let\@evenfoot\@oddfoot}
\makeatother

\def\oto{{\bfig\morphism<180,0>[\mkern-4mu`\mkern-4mu;]\place(86,0)[\circ]\efig}}
\def\rto{{\bfig\morphism<180,0>[\mkern-4mu`\mkern-4mu;]\place(82,0)[\mapstochar]\efig}}

\newcommand{\da}{\downarrow}
\newcommand{\ua}{\uparrow}
\newcommand{\ra}{\multimap}
\newcommand{\la}{\leftarrow}

\newcommand{\bv}{\bigvee}
\newcommand{\bw}{\bigwedge}

\newcommand{\dv}{\dashv}

\newcommand{\al}{\alpha}

\newcommand{\lam}{\lambda}

\newcommand{\Om}{\Omega}

\newcommand{\CA}{\mathcal{A}}

\newcommand{\CC}{\mathcal{C}}

\newcommand{\CK}{\mathcal{K}}

\newcommand{\CM}{\mathcal{M}}

\newcommand{\CQ}{\mathcal{Q}}

\newcommand{\Fix}{{\sf Fix}}

\newcommand{\bbQ}{\mathbb{Q}}
\newcommand{\bbR}{\mathbb{R}}

\newcommand{\bbX}{\mathbb{X}}
\newcommand{\bbY}{\mathbb{Y}}

\newcommand{\Ctx}{{\bf Ctx}}

\newcommand{\Sup}{{\bf Sup}}

\newcommand{\LCtx}{L\text{-}\Ctx}

\newcommand{\LSup}{L\text{-}\Sup}

\newcommand{\aR}{R^{\forall}}
\newcommand{\eR}{R^{\exists}}
\newcommand{\dR}{R^{\da}}
\newcommand{\uR}{R^{\ua}}

\newcommand{\aphi}{\varphi^{\forall}}
\newcommand{\ephi}{\varphi^{\exists}}
\newcommand{\dphi}{\varphi^{\da}}
\newcommand{\uphi}{\varphi^{\ua}}
\newcommand{\apsi}{\psi^{\forall}}
\newcommand{\epsi}{\psi^{\exists}}
\newcommand{\dpsi}{\psi^{\da}}
\newcommand{\upsi}{\psi^{\ua}}

\newcommand{\op}{{\rm op}}

\newcommand{\Mphi}{\CM\varphi}

\newcommand{\Kphi}{\CK\varphi}

\numberwithin{equation}{section}

\allowdisplaybreaks

\begin{document}

\begin{frontmatter}



\title{Reducts of fuzzy contexts: Formal concept analysis vs. rough set theory}


\author{Yuxu Chen}
\ead{yuxuchen1210@sina.com}

\author{Jing Liu}
\ead{921492217@qq.com}

\author{Lili Shen}
\ead{shenlili@scu.edu.cn}

\author{Xiaoye Tang\corref{cor}}
\ead{tang.xiaoye@qq.com}

\cortext[cor]{Corresponding author.}
\address{School of Mathematics, Sichuan University, Chengdu 610064, China}

\begin{abstract}
We postulate the intuitive idea of reducts of fuzzy contexts based on formal concept analysis and rough set theory. For a complete residuated lattice $L$, it is shown that reducts of $L$-contexts in formal concept analysis are interdefinable with reducts of $L$-contexts in rough set theory via negation if, and only if, $L$ satisfies the law of double negation.
\end{abstract}


\end{frontmatter}




\section{Introduction}

Formal concept analysis (FCA) \cite{Ganter1999,Davey2002} and rough set theory (RST) \cite{Pawlak1982,Polkowski2002,Duntsch2002,Yao2004} are effective tools in qualitative data analysis. Both theories are grounded in the common framework of \emph{formal contexts}. Explicitly, a formal context (or \emph{context} for short) is a triple $(X,Y,R)$, where $R\subseteq X\times Y$ is a relation between sets $X$ and $Y$. With $X$ and $Y$ representing the set of \emph{objects} and that of \emph{properties}, respectively, $(x,y)\in R$ is interpreted as the object $x$ having the property $y$. Each context $(X,Y,R)$ induces two pairs of Galois connections between the powersets of $X$ and $Y$,
\[\uR\dv\dR\colon(\mathbf{2}^Y)^{\op}\to \mathbf{2}^X\quad\text{and}\quad\eR\dv\aR\colon \mathbf{2}^Y\to \mathbf{2}^X,\]
and the complete lattices consisting of fixed points of the induced closure operators,
\[\CM R\coloneqq\Fix(\dR\uR)=\{U\subseteq X\mid\dR\uR U=U\}\quad\text{and}\quad\CK R\coloneqq\Fix(\aR\eR)=\{U\subseteq X\mid\aR\eR U=U\},\]
are called the \emph{formal concept lattice} (or, concept lattice based on FCA) and the \emph{property oriented concept lattice} (or, concept lattice based on RST) of $(X,Y,R)$, respectively. It is well known \cite{Yao2004} that 
\[\CM R=\CK(\neg R),\]
where $\neg R=(X\times Y)\setminus R$ is the complement of $R$ in $X\times Y$.

An important problem in the applications of FCA and RST is the reduction of contexts \cite{Ganter1999,Zhang2003a,Mi2004,Zhang2005a,Wang2008,Thangavel2009,Wu2009,Wei2010,Poelmans2014,BenitezCaballero2020,Wasilewski2023,Zhao2023a,Zhang2025}. While handling a large quantity of data, it is often desirable to decrease the size of the set of objects and/or properties without altering the structure of the concept lattice. Intuitively, given a context $(X,Y,R)$, we wish to find subsets $X'\subseteq X$ and $Y'\subseteq Y$, such that 
\[\CM R_{X',Y'}\cong\CM R\quad(\text{or}\ \ \CK R_{X',Y'}\cong\CK R),\]
where $R_{X',Y'}$ is the restriction of $R$ on $X'\times Y'$. Nevertheless, the following example indicates that this intuition requires further clarification:

\begin{exmp}
Let $\bbQ$ be the set of rational numbers. All of the following concept lattices are isomorphic to the Dedekind-MacNeille completion of the partially ordered set $(\bbQ,\leq)$, i.e., the set 
\[\bbR\cup\{-\infty,+\infty\}\]
of extended real numbers equipped with the usual order:
\begin{itemize}
\item the formal concept lattice of the context $(\bbQ,\bbQ,\leq)$,
\item the property oriented concept lattice of the context $(\bbQ,\bbQ,\not\leq)$,
\item the formal concept lattice of the context $(\bbQ\cap[0,1],\bbQ\cap[0,1],\leq)$, and
\item the property oriented concept lattice of the context $(\bbQ\cap[0,1],\bbQ\cap[0,1],\not\leq)$.
\end{itemize}
\end{exmp}

However, it seems questionable whether $(\bbQ\cap[0,1],\bbQ\cap[0,1],\leq)$ can be considered as a reduct of $(\bbQ,\bbQ,\leq)$ in FCA, or whether $(\bbQ\cap[0,1],\bbQ\cap[0,1],\not\leq)$ can be considered as a reduct of $(\bbQ,\bbQ,\not\leq)$ in RST, as restricting $\bbQ$ to $\bbQ\cap[0,1]$ entails a substantial loss of information. Therefore, an appropriate step towards developing a theory of context reduction is to require that $\CM R$ and $\CM R_{X',Y'}$, as well as $\CK R$ and $\CK R_{X',Y'}$, are not only isomorphic, but also isomorphic in a canonical way. 

In order to establish a rigorous theory of reducts of contexts, Shen--Tao--Zhang \cite{Shen2016a} introduced \emph{reducts of $\CQ$-distributors} via \emph{Chu connections} in the framework of \emph{quantaloid-enriched categories}, where $\CQ$ is a small \emph{quantaloid} \cite{Rosenthal1990,Stubbe2005,Stubbe2014}. Explicitly, given a $\CQ$-distributor $\varphi\colon\bbX\oto\bbY$ and $\CQ$-subcategories $\bbX'\subseteq\bbX$, $\bbY'\subseteq\bbY$, four ``comparison $\CQ$-functors'' were constructed between $\Mphi$ and $\Mphi_{\bbX',\bbY'}$, the $\CQ$-categorical version of the formal concept lattices of $(\bbX,\bbY,\varphi)$ and $(\bbX',\bbY',\varphi_{\bbX',\bbY'})$. It was proved that if one of these four $\CQ$-functors is an isomorphism, then so are the other three. Based on this fact, the notion of a reduct of a $\CQ$-distributor in the sense of FCA was postulated, which is fully compatible with the classical notion of \emph{reducibility} of objects and properties presented in \cite{Ganter1999} (see \cite[Remark 4.2.2 and Theorem 4.2.3]{Shen2016a}).

In this paper, we consider reducts of fuzzy contexts valued in a \emph{complete residuated lattice} \cite{Bvelohlavek2002,Hajek1998} 
\[L=(L,*);\] 
that is, reducts of $L$-contexts. The aim of this paper is threefold:
\begin{itemize}
\item We simplify and reformulate the definition of reducts of $L$-contexts in FCA; see Section \ref{Reducts_FCA}. Although a more generalized $\CQ$-categorical version was proposed in \cite{Shen2016a}, we observe that \emph{infomorphisms} (also \emph{Chu transforms}) \cite{Barr1991,Pratt1995,Krotzsch2005,Ganter2007,Shen2013,Shen2013a}, which are a special case of Chu connections, are sufficient for this purpose. This leads to a more concise and accessible formulation.
\item We apply the methodology of Section \ref{Reducts_FCA} to define reducts of $L$-contexts in RST; see Section \ref{Reducts_RST}. Specifically, we construct ``comparison maps'' via Theorem \ref{reduct-RST}, which lead to the definition of reducts of $L$-contexts in RST (see Definition \ref{reduct-RST-def}). Furthermore, we characterize reducts of $L$-contexts in RST in terms of \emph{$\varphi$-reducibility} (see Theorem \ref{RST-reducible-reduct}).
\item We compare reducts of $L$-contexts in FCA and RST in Section \ref{Reducts_FCA_RST}, and show that they are interdefinable via \emph{negation} if, and only if, $L$ satisfies the law of double negation; see Theorem \ref{main}. This is the main result of this paper.
\end{itemize}

The content of this paper is organized as follows. Section \ref{Preliminaries} reviews fundamental concepts of complete residuated lattices, $L$-contexts, $L$-orders and complete $L$-lattices. Sections \ref{Reducts_FCA}--\ref{Reducts_FCA_RST} are devoted to the main theoretical framework and results of this paper as described above. Finally, Sections \ref{Reducts_Examples}--\ref{Reducts_Algorithms} present examples and algorithms to illustrate our results.

\section{$L$-contexts, $L$-orders and complete $L$-lattices} \label{Preliminaries}

Throughout this paper, let 
\[L=(L,*)\]
denote a \emph{complete residuated lattice} \cite{Bvelohlavek2002,Hajek1998} (also known as a \emph{commutative and integral quantale} \cite{Rosenthal1990}). Explicitly:
\begin{itemize}
\item $L$ is a complete lattice with a bottom element $0$ and a top element $1$;
\item $(L,*,1)$ is a commutative monoid;
\item $a*\bv\limits_{i\in I}b_i=\bv\limits_{i\in I}a*b_i$ for all $a,b_i\in L$ $(i\in I)$.
\end{itemize}
Since $(a* -)\colon L\to L$ preserves suprema, there exists a Galois connection
\[(a* -)\dashv(a\ra -)\colon L\to L,\]
where $\ra$ is called the \emph{implication} in $L$ and satisfies
\[a*b\leq c\iff a\leq b\ra c\]
for all $a,b,c\in L$. 


The \emph{negation} on $L$ is the map 
\[\neg\colon L\to L,\quad\neg a=a\ra 0.\] 
$L$ is said to satisfy the \emph{law of double negation} if $\neg\neg=1_L$, the identity map on $L$; that is,
\[\neg\neg a=(a\ra 0)\ra 0=a\]
for all $a\in L$. 

\begin{exmp} \label{L-exmp}
We list here some important complete residuated lattices:
\begin{enumerate}[label=(\arabic*)]
\item \label{L-exmp:frame} Every \emph{frame} $\Om=(\Om,\wedge)$ is a complete residuated lattice, and it satisfies the law of double negation if and only if it is a complete Boolean algebra. 
\item \label{L-exmp:Lawvere} Let $[0,\infty]$ be the extended non-negative real line equipped with the order ``$\geq$'' (so that $0$ becomes the top element and $\infty$ the bottom element). Then $[0,\infty]=([0,\infty],+)$ is a complete residuated lattice, where ``$+$'' is the usual addition extended via
\[p+\infty=\infty+p=\infty\]
to $[0,\infty]$. The implication in $[0,\infty]$ is given by
\[p\ra q=\begin{cases}
q-p & \text{if}\ p<q,\\
0 & \text{else}
\end{cases}\]
for all $p,q\in[0,\infty]$, where the subtraction ``$-$'' is extended via
\[\infty-p=\begin{cases}
\infty & \text{if}\ p<\infty,\\
0 & \text{if}\ p=\infty
\end{cases}\]
to $[0,\infty]$. It is easy to see that $[0,\infty]$ does not satisfy the law of double negation.
\item \label{L-exmp:BL} Every \emph{complete BL-algebra} \cite{Hajek1998} is a complete residuated lattice, and it satisfies the law of double negation if and only if it s a \emph{complete MV-algebra} \cite{Chang1958}. In particular, the unit interval $[0,1]$ equipped with a \emph{continuous t-norm} $*$ \cite{Klement2000,Alsina2006} is a complete residuated lattice\footnote{In fact, for $([0,1],*)$ to be a complete residuated lattice, we only need $*$ to be \emph{left-continuous}.}, and it satisfies the law of double negation if and only if it is isomorphic to $[0,1]$ equipped with the \emph{{\L}ukasiewicz t-norm}. 
\end{enumerate}
\end{exmp}

By an \emph{$L$-relation} we mean a map
\[\varphi\colon X\times Y\to L.\]
We sometimes write such an $L$-relation as $\varphi\colon X\rto Y$, especially in diagrams (e.g. \eqref{XYphi-ABpsi} and \eqref{info-squares}). We denote the \emph{dual} and the \emph{negation} of $\varphi$ by 
\[\varphi^{\op}\colon Y\times X\to L\quad\text{and}\quad\neg\varphi\colon X\times Y\to L,\] 
respectively, with
\[\varphi^{\op}(y,x)=\varphi(x,y)\quad\text{and}\quad(\neg\varphi)(x,y)=\neg\varphi(x,y)\]
for all $x\in X$, $y\in Y$. 

An \emph{$L$-context} is a triple $(X,Y,\varphi)$, where $X$ and $Y$ are (crisp) sets, and $\varphi\colon X\times Y\to L$ is an $L$-relation. With $X$ and $Y$ interpreted as the set of \emph{objects} and that of \emph{properties}, respectively, $\varphi(x,y)$ expresses the degree of the object $x$ having the property $y$.

An \emph{infomorphism} (also \emph{Chu transform}) \cite{Ganter2007,Krotzsch2005,Shen2013}
\[(f,g)\colon(X,Y,\varphi)\to(A,B,\psi)\]
between $L$-contexts is a pair of maps
\[f\colon X\to A\quad\text{and}\quad g\colon B\to Y\]
\begin{equation} \label{XYphi-ABpsi}
\bfig
\square/->`->`->`<-/<600,400>[X`A`Y`B;f`\varphi`\psi`g]
\place(0,200)[-]
\place(600,200)[-]
\efig
\end{equation}
such that
\[\varphi(x,g(b))=\psi(f(x),b)\]
for all $x\in X$, $b\in B$. The category of $L$-contexts and infomorphisms is denoted by
\[\LCtx.\]

An \emph{$L$-order} on a set $X$ is an $L$-relation $\al\colon X\times X\to L$ such that
\begin{itemize}
\item $\al(x,x)=1$,
\item $\al(x,y)*\al(y,z)\leq\al(x,z)$,
\item $\al(x,y)=\al(y,x)=1\implies x=y$
\end{itemize}
for all $x,y,z\in X$. 
We abbreviate an $L$-ordered set $(X,\al)$ to $X$, and write $X(x,y)$ instead of $\al(x,y)$ if no confusion arises. Note that every $L$-ordered set $X$ has a \emph{dual} $X^{\op}$, given by
\[X^{\op}(x,y)=X(y,x)\]
for all $x,y\in X$. The underlying (crisp) order of an $L$-ordered set $X$ is given by 
\[x\leq y\iff X(x,y)=1\]
for all $x,y\in X$. A map $f\colon X\to Y$ between $L$-ordered sets is \emph{monotone} if
\[X(x,x')\leq Y(f(x),f(x'))\] 
for all $x,x'\in X$, and it is \emph{$L$-isometric} (also \emph{fully faithful}) if
\[X(x,x')=Y(f(x),f(x'))\]
for all $x,x'\in X$. Monotone maps are ordered by
\[f\leq g\colon X\to Y\iff \forall x\in X\colon f(x)\leq g(x)\iff\forall x\in X\colon Y(f(x),g(x))=1.\]

Let $X$ be an $L$-ordered set. An \emph{$L$-subset} of a (crisp) set $X$ is a map from $X$ to $L$, and the set of all $L$-subsets on $X$ is denoted by $L^X$, which is equipped with an $L$-order
\[L^X(\mu,\mu')=\bw\limits_{x\in X}\mu(x)\ra\mu'(x)\]
for all $\mu,\mu'\in L^X$. For each $\mu\in L^X$, an element $a\in X$ is a \emph{supremum} (resp. \emph{infimum}) of $\mu$ if
\[X(a,x)=\bw\limits_{y\in X}\mu(y)\ra X(y,x)\quad\Big(\text{resp.}\ X(x,a)=\bw\limits_{y\in X}\mu(y)\ra X(x,y)\Big)\]
for all $x\in X$. 
$X$ is called a \emph{complete $L$-lattice} if every $\mu\in L^X$ has a supremum; or equivalently, if every $\mu\in L^X$ has an infimum. In particular, $L^X$ is a complete $L$-lattice.

A pair of monotone maps $f\colon X\to Y$ and $g\colon Y\to X$ between $L$-ordered sets forms an \emph{$L$-adjunction}, denoted by $f\dv g$, if 
\[Y(f(x),y)=X(x,g(y))\]
for all $x\in X$, $y\in Y$. In this case, $f$ is called a \emph{left adjoint} of $g$, and $g$ is a \emph{right adjoint} of $f$, and it necessarily holds that
\begin{equation} \label{ad}
f\circ g\circ f=f\quad\text{and}\quad g\circ f\circ g=g.
\end{equation}
The category of complete $L$-lattices and left adjoints is denoted by
\[\LSup.\]
In particular, every map $f\colon X\to Y$ between (crisp) sets induces an $L$-adjunction $f^{\rightarrow}\dv f^{\la}\colon L^Y\to L^X$, given by
\[f^{\rightarrow}\colon L^X\to L^Y,\quad f^{\rightarrow}(\mu)(y)=\bv\limits_{y=f(x)}\mu(x),\]
\[f^{\la}\colon L^Y\to L^X,\quad f^{\la}(\lam)(x)=\lam(f(x)).\]



An \emph{$L$-closure operator} on an $L$-ordered set $X$ is a monotone map $c\colon X\to X$ such that
\[1_X\leq c\quad\text{and}\quad cc=c.\]
It is easy to see that every $L$-adjunction $f\dv g\colon Y\to X$ induces an $L$-closure operator
\[gf\colon X\to X.\]

\begin{prop} \label{Fix} (See \cite{Shen2013,Shen2013a}.)
Let $c$ be an $L$-closure operator on an $L$-ordered set $X$. Let
\[\Fix(c)\coloneqq\{x\in X\mid c(x)=x\}=\{c(x)\mid x\in X\}.\]
Then
\begin{enumerate}[label={\rm(\arabic*)}]
\item \label{fix-ad} the inclusion map $\Fix(c)\ \to/^(->/ X$ is right adjoint to the codomain restriction $c\colon X\to\Fix(c)$, and
\item \label{fix-cl} $\Fix(c)$ is a complete $L$-lattice provided so is $X$.
\end{enumerate}
\end{prop}

An \emph{$L$-closure space} $(X,c)$ consists of a set $X$ and an $L$-closure operator $c\colon L^X\to L^X$. A map 
\[f\colon(X,c)\to(Y,d)\]
between $L$-closure spaces is \emph{continuous} if $f^{\la}(\lam)\in\Fix(c)$ whenever $\lam\in\Fix(d)$.

\begin{prop} \label{Fix-adjoint} (See \cite{Shen2013,Shen2013a}.)
If $f\colon(X,c)\to(Y,d)$ is a continuous map between $L$-closure spaces, then 
\[f^{\triangleright}\dv f^{\triangleleft}\colon\Fix(d)\to\Fix(c),\] 
where
\[f^{\triangleright}(\mu)=df^{\rightarrow}(\mu)\quad\text{and}\quad f^{\triangleleft}(\lam)=f^{\la}(\lam)\]
for all $\mu\in\Fix(c)$, $\lam\in\Fix(d)$.
\end{prop}

\section{Reducts of $L$-contexts in formal concept analysis} \label{Reducts_FCA}

Each $L$-context $(X,Y,\varphi)$ induces an $L$-adjunction $\uphi\dv\dphi$, given by
\begin{align*}
\uphi\colon & L^X\to (L^Y)^\op,\quad \uphi(\mu)(y)=\bw\limits_{x\in X}\mu(x)\ra\varphi(x,y),\\
\dphi\colon & (L^Y)^\op\to L^X,\quad \dphi(\lambda)(x)=\bw\limits_{y\in Y}\lambda(y)\ra\varphi(x,y).
\end{align*}
We denote by
\[\Mphi\coloneqq\Fix(\dphi\uphi)=\{\mu\in L^X\mid\dphi\uphi(\mu)=\mu\}\]
the \emph{formal concept $L$-lattice} of $(X,Y,\varphi)$ (or, the concept $L$-lattice of $(X,Y,\varphi)$ based on FCA), which consists of fixed points of the $L$-closure operator $\dphi\uphi$, and it is a complete $L$-lattice by Proposition \ref{Fix}\ref{fix-cl}. 
%
%
It is well known \cite{Shen2013,Shen2013a,Shen2014,Shen2016a} that mapping each $L$-context $(X,Y,\varphi)$ to its formal concept $L$-lattice $\CM\varphi$ yields a functor
\[\CM\colon\LCtx\to\LSup.\]
Explicitly, each infomorphism
\[(f,g)\colon(X,Y,\varphi)\to(A,B,\psi)\]
between $L$-contexts induces a continuous map
\[f\colon(X,\dphi\uphi)\to(Y,\dpsi\upsi)\]
between $L$-closure spaces (see \cite[Proposition 5.1]{Shen2013}); and consequently, 
\begin{equation} \label{M(f,g)-def}
\CM(f,g)\coloneqq f^{\triangleright}=(\CM\varphi\ \to/^(->/ L^X\to^{f^{\rightarrow}} L^A\to^{\psi^\da\psi^\ua}\CM\psi)
\end{equation}
is a left adjoint between formal concept $L$-lattices (see Proposition \ref{Fix-adjoint}), whose right adjoint is given by 
\begin{equation} \label{M(f,g)*-def}
\CM(f,g)^*\coloneqq f^{\triangleleft}\colon\CM\psi\to\CM\varphi.
\end{equation}

For each (crisp) set $X$, we denote by $X'$ a (crisp) subset of $X$. For each $L$-context $(X,Y,\varphi)$ and $X'\subseteq X$, $Y'\subseteq Y$, we write 
\[\tau\colon X'\to X\quad\text{and}\quad \nu\colon Y'\to Y\]
for the inclusion maps, and 
\[\varphi_{X',Y'}\colon X'\times Y'\to L\]
for the restriction of $\varphi\colon X\times Y\to L$ on $X'$ and $Y'$. The triple $(X',Y',\varphi_{X',Y'})$ is thus an \emph{$L$-subcontext} of $(X,Y,\varphi)$. In particular, we write $\mu_{X'}$ for the restriction of $\mu\in L^X$ on $X'$. Conversely, for an $L$-relation $\psi\colon X'\times Y'\to L$, we define
\[\underline{\psi}\colon X\times Y\to L,\quad \underline{\psi}(x,y)=\begin{cases}
\psi(x,y) &\text{if}\ x\in X',\ y\in Y',\\
0 &\text{else.}
\end{cases}\]
for the extension of $\psi$ to $X$ and $Y$.

For each $L$-context $(X,Y,\varphi)$ and $X'\subseteq X$, $Y'\subseteq Y$, it is easy to see that
\[(\tau,1_Y)\colon \varphi_{X',Y}\to\varphi\quad\text{and}\quad(1_X,\nu)\colon \varphi\to\varphi_{X,Y'}\]
\begin{equation} \label{info-squares}
\bfig
\square(-600,0)/->`->`->`<-/<600,400>[X'`X`Y`Y;\tau`\varphi_{X',Y}`\varphi`1_Y]
\place(-600,200)[-]
\place(0,200)[-]
\square/->`->`->`<-/<600,400>[X`X`Y`Y';1_X``\varphi_{X,Y'}`\nu]
\place(0,200)[-]
\place(600,200)[-]
\efig
\end{equation}
are infomorphisms. Thus we obtain left adjoints:
\begin{equation} \label{M(tau,1_Y)}
\CM\varphi_{X',Y}\to^{\CM(\tau,1_Y)}\CM\varphi\to^{\CM(1_X,\nu)}\CM\varphi_{X,Y'}.
\end{equation}
Replacing $X$ by $X'$ in the left square of \eqref{info-squares} and $Y$ by $Y'$ in the right square of \eqref{info-squares} we obtain another two left adjoints:
\begin{equation} \label{M(1_{X'},J)}
\CM\varphi_{X',Y}\to^{\CM(1_{X'},\nu)}\CM\varphi_{X',Y'}\to^{\CM(\tau,1_{Y'})}\CM\varphi_{X,Y'}.
\end{equation}
From Proposition \ref{Fix-adjoint}, \eqref{M(f,g)-def}, \eqref{M(f,g)*-def}, \eqref{M(tau,1_Y)} and \eqref{M(1_{X'},J)} we derive four canonical maps between $\CM\varphi$ and $\CM\varphi_{X',Y'}$:
\begin{equation} \label{canonical-maps-M-1}
\bfig
\square|blrb|/@{->}@<-5pt>`->`->`@{->}@<-5pt>/<1200,500>[\CM\varphi`\CM\varphi_{X',Y}`\CM\varphi_{X,Y'}`\CM\varphi_{X',Y'};\CM(\tau,1_Y)^*`\CM(1_X,\nu)`\CM(1_{X'},\nu)`\CM(\tau,1_{Y'})^*]
\morphism(0,500)|a|/@{<-}@<5pt>/<1200,0>[\CM\varphi`\CM\varphi_{X',Y};\CM(\tau,1_Y)]
\morphism|a|/@{<-}@<5pt>/<1200,0>[\CM\varphi_{X,Y'}`\CM\varphi_{X',Y'};\CM(\tau,1_{Y'})]
\place(600,0)[\bot]
\place(600,500)[\bot]
\efig
\end{equation}
\begin{equation} \label{canonical-maps-M-2}
\bfig
\square|blrb|/@{->}@<-5pt>`->`->`@{->}@<-5pt>/<1200,500>[\CM\varphi_{X',Y'}`\CM\varphi_{X',Y}`\CM\varphi_{X,Y'}`\CM\varphi;\CM(1_{X'},\nu)^*`\CM(\tau,1_{Y'})`\CM(\tau,1_Y)`\CM(1_X,\nu)^*]
\morphism(0,500)|a|/@{<-}@<5pt>/<1200,0>[\CM\varphi_{X',Y'}`\CM\varphi_{X',Y};\CM(1_{X'},\nu)]
\morphism|a|/@{<-}@<5pt>/<1200,0>[\CM\varphi_{X,Y'}`\CM\varphi;\CM(1_X,\nu)]
\place(600,0)[\bot]
\place(600,500)[\bot]
\efig
\end{equation}
\begin{itemize}
\item the right-down route of \eqref{canonical-maps-M-1} gives the map
\begin{equation} \label{R1-def}
R_1\coloneqq\left(\CM\varphi\to^{\CM(\tau,1_Y)^*}\CM\varphi_{X',Y}\to^{\CM(1_{X'},\nu)}\CM\varphi_{X',Y'}\right),\quad R_1(\mu)=(\varphi_{X',Y'})^\da(\varphi_{X',Y'})^\ua(\mu_{X'});
\end{equation}
\item the down-right route of \eqref{canonical-maps-M-1} gives the map
\begin{equation} \label{R2-def}
R_2\coloneqq\left(\CM\varphi\to^{\CM(1_X,\nu)}\CM\varphi_{X,Y'}\to^{\CM(\tau,1_{Y'})^*}\CM\varphi_{X',Y'}\right),\quad R_2(\mu)=\left((\varphi_{X,Y'})^\da(\varphi_{X,Y'})^\ua(\mu)\right)_{X'};
\end{equation}
\item the right-down route of \eqref{canonical-maps-M-2} gives the map
\begin{equation} \label{E1-def}
E_1\coloneqq\left(\CM\varphi_{X',Y'}\to^{\CM(1_{X'},\nu)^*}\CM\varphi_{X',Y}\to^{\CM(\tau,1_Y)} \CM\varphi\right),\quad E_1(\mu')=\dphi\uphi(\underline{\mu'});
\end{equation}
\item the down-right route of \eqref{canonical-maps-M-2} gives the map
\begin{equation} \label{E2-def}
E_2\coloneqq\left(\CM\varphi_{X'Y'}\to^{\CM(\tau,1_{Y'})}\CM\varphi_{X,Y'}\to^{\CM(1_X,\nu)^*}\CM\varphi\right),\quad E_2(\mu')=(\varphi_{X,Y'})^\da(\varphi_{X,Y'})^\ua(\underline{\mu'}).\
\end{equation}
\end{itemize}

\begin{rem}
The expressions of the maps $R_1$, $E_1$, $R_2$ and $E_2$ given in \eqref{R1-def}--\eqref{E2-def} are a special case of those in \cite[Subsection 4.1]{Shen2016a}, and they can be derived in a similar manner as in the next section for the maps $S_1$, $F_1$, $S_2$ and $F_2$ (see \eqref{S1-def}--\eqref{F2-def}). So, the details are omitted here.
\end{rem}

\begin{thm} (See \cite[Theorem 4.1.1]{Shen2016a}.)
For each $L$-context $(X,Y,\varphi)$ and $X'\subseteq X$, $Y'\subseteq Y$, the following statements are equivalent:
\begin{enumerate}[label={\rm(\roman*)}]
\item $R_1$ is an isomorphism of complete $L$-lattices;
\item $R_2$ is an isomorphism of complete $L$-lattices;
\item $E_1$ is an isomorphism of complete $L$-lattices;
\item $E_2$ is an isomorphism of complete $L$-lattices.
\end{enumerate}
\end{thm}

The above theorem indicates that each of $R_1$, $R_2$, $E_1$ and $E_2$ may act as a ``comparison map'' between $\Mphi$ and $\Mphi_{X',Y'}$. Therefore, it is natural to define reducts of $L$-contexts in FCA as follows:

\begin{defn} \label{reduct-FCA} (See \cite[Definition 4.1.2]{Shen2016a}.)
Given an $L$-context $(X,Y,\varphi)$ and $X'\subseteq X$, $Y'\subseteq Y$, we say that $(X',Y',\varphi_{X',Y'})$ is a \emph{reduct} of $(X,Y,\varphi)$ in FCA if any one (hence each) of $R_1$, $R_2$, $E_1$, $E_2$ is an isomorphism.
\end{defn}

For each $X'\subseteq X$, we write 
\[X\setminus X'=\{x\in X\mid x\not\in X'\}\] 
for the complement of $X'$ in $X$.  We may also characterize reducts of $L$-contexts in terms of reducible subsets:

\begin{defn} \label{phi-reducible} (See \cite[Definition 4.2.1]{Shen2016a}.)
Given an $L$-context $(X,Y,\varphi)$ and $X'\subseteq X$, we say that $X\setminus X'$ is \emph{$\varphi$-reducible} in FCA if for any $\mu\in L^X$, there exits $\mu'\in L^{X'}$ such that
\begin{equation}
\varphi^\uparrow(\mu)=(\varphi_{X',Y})^\uparrow(\mu'). 
\end{equation}
Dually, for $Y'\subseteq Y$, we say that $Y\setminus Y'$ is \emph{$\varphi$-reducible} in FCA if for any $\lambda\in (L^Y)^\op$, there exists $\lambda'\in (L^{Y'})^\op$ such that
\begin{equation}
\dphi(\lambda)=(\varphi_{X,Y'})^\da(\lambda').
\end{equation}
\end{defn}

\begin{prop} \label{FCA-reducible-X} (See \cite[Propositions 4.2.4 and 4.2.5]{Shen2016a}.)
Let $(X,Y,\varphi)$ be an $L$-context and $X'\subseteq X$, $Y'\subseteq Y$. Then
\begin{enumerate}[label={\rm(\arabic*)}]
\item \label{FCA-reducible-X:X} $X\setminus X'$ is $\varphi$-reducible in FCA if, and only if,
\[\uphi\dphi=(\varphi_{X',Y})^\ua(\varphi_{X',Y})^\da\colon (L^Y)^\op\to(L^Y)^\op.\]
\item \label{FCA-reducible-X:Y} $Y\setminus Y'$ is $\varphi$-reducible in FCA if, and only if,
\[\dphi\uphi=(\varphi_{X,Y'})^\da(\varphi_{X,Y'})^\ua\colon L^X\to L^X.\]
\end{enumerate}
\end{prop}

\begin{thm} \label{FCA-reducible-reduct} (See \cite[Theorem 4.2.3]{Shen2016a}.) 
An $L$-context $(X',Y',\varphi_{X',Y'})$ is a reduct of $(X,Y,\varphi)$ in FCA if, and only if, both $X\setminus X'$ and $Y\setminus Y'$ are $\varphi$-reducible in FCA.
\end{thm}

\begin{rem} \label{reducible}
Let $(X,Y,R)$ be a (crisp) context; that is, $R\subseteq X\times Y$ is a (crisp) relation. In formal concept analysis \cite{Ganter1999}, an object $x\in X$ is said to be \emph{reducible} if 
\[\uR(\{x\})=\uR(U)\]
for some $U\subseteq X\setminus\{x\}$, and a property $y\in Y$ is said to be \emph{reducible} if
\[\dR(\{y\})=\dR(V)\]
for some $V\subseteq Y\setminus\{y\}$. For each $X'\subseteq X$ and $Y'\subseteq Y$, it is straightforward to check the following facts:
\begin{itemize}
\item $X\setminus X'$ is $R$-reducible in the sense of Definition \ref{phi-reducible} if, and only if, every $x\in X\setminus X'$ is reducible;
\item $Y\setminus Y'$ is $R$-reducible in the sense of Definition \ref{phi-reducible} if, and only if, every $y\in Y\setminus Y'$ is reducible.
\end{itemize}
Therefore, the \emph{$\varphi$-reducibility} introduced in Definition \ref{phi-reducible} is compatible with the classical notion of reducibility in formal concept analysis. Moreover, Theorem \ref{FCA-reducible-reduct} justifies the notion of \emph{reduct} in FCA introduced in Definition \ref{reduct-FCA}.
\end{rem}

In the next section, we will establish the above results in rough set theory.

\section{Reducts of $L$-contexts in rough set theory} \label{Reducts_RST}

Each $L$-context $(X,Y,\varphi)$ induces another $L$-adjunction $\ephi\dv\aphi$, given by
\begin{align*}
\ephi\colon & L^X\to L^Y,\quad\ephi(\mu)(y)=\bv\limits_{x\in X}\mu(x)*\varphi(x,y),\\
\aphi\colon & L^Y\to L^X,\quad\aphi(\lambda)(x)=\bw\limits_{y\in Y}\varphi(x,y)\ra\lambda(y).
\end{align*}
We denote by
\[\Kphi\coloneqq\Fix(\aphi\ephi)=\{\mu\in L^X\mid\aphi\ephi(\mu)=\mu\}\]
the \emph{property oriented concept $L$-lattice} of $(X,Y,\varphi)$ (or, the concept $L$-lattice of $(X,Y,\varphi)$ based on RST), which consists of fixed points of the $L$-closure operator $\aphi\ephi$, and it is also a complete $L$-lattice by Proposition \ref{Fix}\ref{fix-cl}. 
Mapping an $L$-context $(X,Y,\varphi)$ to its property oriented concept $L$-lattice $\CK\varphi$ also gives rise to a functor 
\[\CK\colon\LCtx\to\LSup.\]
Explicitly, each infomorphism
\[(f,g)\colon(X,Y,\varphi)\to(A,B,\psi)\]
between $L$-contexts induces a continuous map
\[f\colon(X,\aphi\ephi)\to(Y,\apsi\epsi)\]
between $L$-closure spaces (see \cite[Proposition 6.4]{Shen2013}); and consequently, 
\begin{equation} \label{K(f,g)-def}
\CK(f,g)\coloneqq f^{\triangleright}=(\CK\varphi\ \to/^(->/ L^X\to^{f^{\rightarrow}} L^A\to^{\aphi\ephi}\CK\psi)
\end{equation}
is a left adjoint between property oriented concept $L$-lattices (Proposition \ref{Fix-adjoint}), whose right adjoint is given by 
\begin{equation} \label{K(f,g)*-def}
\CK(f,g)^*\coloneqq f^{\triangleleft}\colon\CK\psi\to\CK\varphi.
\end{equation} 

Given an $L$-context $(X,Y,\varphi)$ and (crisp) subsets $X'\subseteq X$, $Y'\subseteq Y$, similarly to \eqref{canonical-maps-M-1} and \eqref{canonical-maps-M-2} we may construct four canonical maps between $\Kphi$ and $\Kphi_{X',Y'}$:
\begin{equation} \label{canonical-maps-K-1}
\bfig
\square|blrb|/@{->}@<-5pt>`->`->`@{->}@<-5pt>/<1200,500>[\CK\varphi`\CK\varphi_{X',Y}`\CK\varphi_{X,Y'}`\CK\varphi_{X',Y'};\CK(\tau,1_Y)^*`\CK(1_X,\nu)`\CK(1_{X'},\nu)`\CK(\tau,1_{Y'})^*]
\morphism(0,500)|a|/@{<-}@<5pt>/<1200,0>[\CK\varphi`\CK\varphi_{X',Y};\CK(\tau,1_Y)]
\morphism|a|/@{<-}@<5pt>/<1200,0>[\CK\varphi_{X,Y'}`\CK\varphi_{X',Y'};\CK(\tau,1_{Y'})]
\place(600,0)[\bot]
\place(600,500)[\bot]
\efig
\end{equation}
\begin{equation} \label{canonical-maps-K-2}
\bfig
\square|blrb|/@{->}@<-5pt>`->`->`@{->}@<-5pt>/<1200,500>[\CK\varphi_{X',Y'}`\CK\varphi_{X',Y}`\CK\varphi_{X,Y'}`\CK\varphi;\CK(1_{X'},\nu)^*`\CK(\tau,1_{Y'})`\CK(\tau,1_Y)`\CK(1_X,\nu)^*]
\morphism(0,500)|a|/@{<-}@<5pt>/<1200,0>[\CK\varphi_{X',Y'}`\CK\varphi_{X',Y};\CK(1_{X'},\nu)]
\morphism|a|/@{<-}@<5pt>/<1200,0>[\CK\varphi_{X,Y'}`\CK\varphi;\CK(1_X,\nu)]
\place(600,0)[\bot]
\place(600,500)[\bot]
\efig
\end{equation}
\begin{itemize}
\item the right-down route of \eqref{canonical-maps-K-1} gives the map
\[S_1\coloneqq\left(\CK\varphi\to^{\CK(\tau,1_Y)^*}\CK\varphi_{X',Y}\to^{\CK(1_{X'},\nu)}\CK\varphi_{X',Y'}\right);\]
\item the down-right route of \eqref{canonical-maps-K-1} gives the map
\[S_2\coloneqq\left(\CK\varphi\to^{\CK(1_X,\nu)}\CK\varphi_{X,Y'}\to^{\CK(\tau,1_{Y'})^*}\CK\varphi_{X',Y'}\right);\]
\item the right-down route of \eqref{canonical-maps-K-2} gives the map
\[F_1\coloneqq\left(\CK\varphi_{X',Y'}\to^{\CK(1_{X'},\nu)^*}\CK\varphi_{X',Y}\to^{\CK(\tau,1_Y)} \CK\varphi\right);\]
\item the down-right route of \eqref{canonical-maps-K-2} gives the map
\[F_2\coloneqq\left(\CK\varphi_{X'Y'}\to^{\CK(\tau,1_{Y'})}\CK\varphi_{X,Y'}\to^{\CK(1_X,\nu)^*}\CK\varphi\right).\]
\end{itemize}

To compute the maps $S_1$, $S_2$, $F_1$ and $F_2$, we present the following lemmas:

\begin{lem} \label{cd}
\begin{enumerate}[label={\rm(\arabic*)}]
\item \label{restric} $(\varphi_{X,Y'})^\exists(\mu)=\left(\ephi(\mu)\right)_{Y'}$ and $\left(\aphi(\lam)\right)_{X'}$ = $(\varphi_{X',Y})^\forall(\lam)$ for all $\mu\in L^X$, $\lam\in L^Y$.
\item \label{underline} $(\varphi_{X',Y})^\exists(\mu')=\ephi(\underline{\mu'})$ for all $\mu'\in L^{X'}$.
\end{enumerate}
\end{lem}

\begin{proof}
Straightforward calculations.
\end{proof}

\begin{lem} \label{lem}
\begin{enumerate}[label={\rm(\arabic*)}]
\item \label{lem-1} $\CK\varphi_{X,Y'}\subseteq\CK\varphi$, and $\CK(1_X,\nu)^*\colon \CK\varphi_{X,Y'}\ \to/^(->/\CK\varphi$ is the inclusion map.
\item \label{lem-2} $\CK(\tau,1_Y)\colon\CK\varphi_{X',Y}\to\CK\varphi$ is $L$-isometric, and $\CK(\tau,1_Y)^*\colon\CK\varphi\to\CK\varphi_{X',Y}$ sends each $\mu\in\CK\varphi$ to the restriction $\mu_{X'}\in \CK\varphi_{X',Y}$.
\end{enumerate}
\end{lem}

\begin{proof}
\ref{lem-1} By Proposition \ref{Fix-adjoint} and Equation \eqref{K(f,g)*-def}, 
\[\CK(1_X,\nu)^*(\mu)=1_X^{\la}(\mu)=\mu\]
for all $\mu\in\CK\varphi_{X,Y'}$. The conclusion thus follows.


\ref{lem-2} First, it follows from Proposition \ref{Fix-adjoint} and Equation \eqref{K(f,g)*-def} that
\[\CK(\tau,1_Y)^*(\mu)(x')=\tau^{\la}(\mu)(x')=\mu(\tau(x'))=\mu(x')=\mu_{X'}(x')\]
for all $\mu\in\Kphi$, $x'\in X'$. Thus $\CK(\tau,1_Y)^*=(-)_{X'}$.

Second, for any $\mu'\in\CK\varphi_{X',Y}$, since
\[\tau^{\rightarrow}(\mu')(x)=\bv_{x=\tau(x')}\mu'(x')=\begin{cases}
\mu'(x) & \text{if}\ x\in X'\\
0 & \text{else}
\end{cases}=\underline{\mu'}(x)\]
for all $x\in X$, by Equation \eqref{K(f,g)-def} we have
\[\CK(\tau,1_Y)(\mu')=\aphi\ephi\tau^{\rightarrow}(\mu')=\aphi\ephi(\underline{\mu'}).\]
Thus
\begin{align*}
L^{X'}(\mu',\mu'')&=L^{X'}\left(\mu',(\varphi_{X',Y})^\forall(\varphi_{X',Y})^\exists(\mu'')\right)&(\mu''\in\CK\varphi_{X',Y})\\
&=L^Y\left((\varphi_{X',Y})^\exists(\mu'),(\varphi_{X',Y})^\exists(\mu'')\right)&((\varphi_{X',Y})^\exists\dv(\varphi_{X',Y})^\forall)\\
&=L^Y\left(\ephi(\underline{\mu'}),\ephi(\underline{\mu''})\right) &(\text{Lemma \ref{cd}\ref{underline}})\\
&=L^Y\left(\ephi\aphi\ephi(\underline{\mu'}),\ephi(\underline{\mu''})\right)&(\text{Equation \eqref{ad}})\\
&=L^X\left(\aphi\ephi(\underline{\mu'}),\aphi\ephi(\underline{\mu''})\right)&(\ephi\dv\aphi)\\
&=L^X\Big(\CK(\tau,1_Y)(\mu'),\CK(\tau,1_Y)(\mu'')\Big)
\end{align*}
for all $\mu',\mu''\in\CK\varphi_{X',Y}$, indicating that $\CK(\tau,1_Y)$ is $L$-isometric.
\end{proof}

Therefore, the maps $S_1$, $S_2$, $F_1$ and $F_2$ become:
\begin{align}
&S_1=\left(\CK\varphi\to^{(-)_{X'}}\CK\varphi_{X',Y}\to^{\CK(1_X',J)}\CK\varphi_{X',Y'}\right),\quad S_1(\mu)=(\varphi_{X',Y'})^\forall(\varphi_{X',Y'})^\exists(\mu_{X'}), \label{S1-def}\\
&F_1=\left(\CK\varphi_{X',Y'}\ \to/^(->/\CK\varphi_{X',Y}\to^{\CK(\tau,1_Y)}\CK\varphi\right),\quad\quad\   F_1(\mu')=\aphi\ephi(\underline{\mu'}), \label{S2-def}\\
&S_2=\left(\CK\varphi\to^{\CK(1_X,\nu)}\CK\varphi_{X,Y'}\to^{(-)_{X'}}\CK\varphi_{X',Y'}\right),\quad S_2(\mu)=\left((\varphi_{X,Y'})^\forall(\varphi_{X,Y'})^\exists(\mu)\right)_{X'}, \label{F1-def}\\
&F_2=\left(\CK\varphi_{X',Y'}\to^{\CK(\tau,1_{Y'})}\CK\varphi_{X,Y'}\ \to/^(->/\CK\varphi\right),\quad\quad\   F_2(\mu')=(\varphi_{X,Y'})^\forall(\varphi_{X,Y'})^\exists(\underline{\mu'}). \label{F2-def}
\end{align}

\begin{lem} \label{SF-id}
\begin{enumerate}[label={\rm(\arabic*)}]
\item \label{SF-id:d} $(F_1\mu')_{X'}=(F_2\mu')_{X'}$ for all $\mu'\in\Kphi_{X',Y'}$.
\item \label{SF-id:id} All of the composites $S_1 F_1$, $S_1 F_2$, $S_2 F_1$, $S_2 F_2$ coincide with the identity map on $\CK\varphi_{X',Y'}$.
\end{enumerate}
\end{lem}

\begin{proof}
\ref{SF-id:d} By Lemma \ref{cd}, 
\begin{equation} \label{equation-restric}
\left(\aphi\ephi(\underline{\mu'})\right)_{X'}=(\varphi_{X',Y})^\forall\ephi(\underline{\mu'})=(\varphi_{X',Y})^\forall(\varphi_{X',Y})^\exists(\mu')=\mu'
\end{equation}
for all $\mu'\in\CK\varphi_{X',Y}$. Thus, for any $\mu'\in\CK\varphi_{X',Y'}\subseteq\CK\varphi_{X',Y}$ (cf. Lemma \ref{lem}\ref{lem-1}), 
\[(F_1(\mu'))_{X'}=\left(\aphi\ephi(\underline{\mu'})\right)_{X'}=\mu'=\left((\varphi_{X,Y'})^\forall(\varphi_{X,Y'})^\exists(\underline{\mu'})\right)_{X'}=(F_2(\mu'))_{X'},\]
where the third equality follows by applying Equation \eqref{equation-restric} to $\varphi_{X,Y'}$.

\ref{SF-id:id} Let $\mu'\in\CK\varphi_{X',Y'}$. Note that
\begin{align*}
S_1 F_1(\mu')&= (\varphi_{X',Y'})^\forall(\varphi_{X',Y'})^\exists\left(\left(\aphi\ephi(\underline{\mu'})\right)_{X'}\right)\\
&=(\varphi_{X',Y'})^\forall(\varphi_{X',Y'})^\exists(\mu')&(\text{Equation \eqref{equation-restric}})\\
&=\mu'&(\mu'\in\CK\varphi_{X',Y'})\\
&=\left((\varphi_{X,Y'})^\forall(\varphi_{X,Y'})^\exists(\underline{\mu'})\right)_{X'}&(\text{Equation \eqref{equation-restric}})\\
&=\left((\varphi_{X,Y'})^\forall(\varphi_{X,Y'})^\exists(\varphi_{X,Y'})^\forall(\varphi_{X,Y'})^\exists(\underline{\mu'})\right)_{X'}&((\varphi_{X,Y'})^\exists\dv(\varphi_{X,Y'})^\forall)\\
&=S_2 F_2(\mu')
\end{align*}
and
\begin{align*}
S_1 F_2(\mu')&= \left(\varphi_{X',Y'})^\forall(\varphi_{X',Y'})^\exists((\varphi_{X,Y'})^\forall(\varphi_{X,Y'})^\exists(\underline{\mu'})\right)_{X'}\\
&=(\varphi_{X',Y'})^\forall(\varphi_{X',Y'})^\exists(\mu')&(\text{Equation \eqref{equation-restric}})\\
&=\mu'&(\mu'\in\CK\varphi_{X',Y'})\\
&=\left((\varphi_{X,Y'})^\forall(\varphi_{X,Y'})^\exists(\underline{\mu'})\right)_{X'}&(\text{Equation \eqref{equation-restric}})\\
&=\left((\varphi_{X,Y'})^\forall\left(\ephi(\underline{\mu'})\right)_{Y'}\right)_{X'}&(\text{Lemma \ref{cd}\ref{restric}})\\
&=\left((\varphi_{X,Y'})^\forall\left(\ephi\aphi\ephi(\underline{\mu'})\right)_{Y'}\right)_{X'}&(\ephi\dv\aphi)\\
&=\left((\varphi_{X,Y'})^\forall(\varphi_{X,Y'})^\exists\aphi\ephi(\underline{\mu'})\right)_{X'}&(\text{Lemma \ref{cd}\ref{restric}})\\
&=S_2 F_1(\mu').
\end{align*}
Hence, all of $S_1 F_1$, $S_1 R_2$, $S_2 F_1$, $S_2 F_2$ coincide with the identity map on $\CK\varphi_{X',Y'}$. 
\end{proof}

\begin{thm} \label{reduct-RST}
For each $L$-context $(X,Y,\varphi)$ and $X'\subseteq X$, $Y'\subseteq Y$, the following statements are equivalent:
\begin{enumerate}[label={\rm(\roman*)}]
\item $S_1$ is an isomorphism of complete $L$-lattices;
\item $S_2$ is an isomorphism of complete $L$-lattices;
\item $F_1$ is an isomorphism of complete $L$-lattices;
\item $F_2$ is an isomorphism of complete $L$-lattices.
\end{enumerate}
\end{thm}

\begin{proof}
As each of the compositions $S_1 F_1$, $S_1 F_2$, $S_2 F_1$ and $S_2 F_2$ coincides with the identity map on $\CK\varphi_{X',Y'}$, it follows that each one of $S_1$, $S_2$, $F_1$ and $F_2$ is an isomorphism if and only if any one of them is. 
\end{proof}

The above theorem allows us to treat $S_1$, $S_2$, $F_1$ and $F_2$ as ``comparison maps'' between $\Kphi$ and $\Kphi_{X',Y'}$. Therefore, reducts of $L$-contexts in RST can be defined as follows:

\begin{defn} \label{reduct-RST-def}
Given an $L$-context $(X,Y,\varphi)$ and $X'\subseteq X$, $Y'\subseteq Y$, we say that $(X',Y',\varphi_{X',Y'})$ is a reduct of $(X,Y,\varphi)$ in RST if any one (hence each) of $S_1$, $S_2$, $F_1$, $F_2$ is an isomorphism.
\end{defn} 

Reducts of $L$-contexts in RST can also be characterized through reducible subsets:

\begin{defn}
Let $(X,Y,\varphi)$ be an $L$-context and $X'\subseteq X$. $X\setminus X'$ is \emph{$\varphi$-reducible} in RST if for any $\mu\in L^X$, there exists $\mu'\in L^{X'}$ such that
\begin{equation}
\ephi(\mu)=(\varphi_{X',Y})^\exists(\mu'). 
\end{equation}
Dually, for $Y'\subseteq Y$, $Y\setminus Y'$ is \emph{$\varphi$-reducible} in RST if for any $\lambda\in L^Y$, there exists $\lambda'\in L^{Y'}$ such that
\begin{equation}
\aphi(\lambda)=(\varphi_{X,Y'})^\forall(\lambda').
\end{equation}
\end{defn}

\begin{prop} \label{RST-reducible-X}
Let $(X,Y,\varphi)$ be an $L$-context and $X'\subseteq X$. The following statements are equivalent:
\begin{enumerate}[label={\rm(\roman*)}]
\item \label{RST-reducible-X:r} $X\setminus X'$ is $\varphi$-reducible in RST.
\item \label{RST-reducible-X:e} $\ephi(\mu)=(\varphi_{X',Y})^\exists\left(\aphi\ephi(\mu)\right)_{X'}$ for all $\mu\in L^X$.

\item \label{RST-reducible-X:a} $\ephi\aphi=(\varphi_{X',Y})^\exists(\varphi_{X',Y})^\forall\colon L^Y\to L^Y$.
\item \label{RST-reducible-X:s} $\CK(\tau,1_Y)\colon\CK\varphi_{X',Y}\to\CK\varphi$ is surjective, thus an isomorphism.
\end{enumerate}
\end{prop}

\begin{proof} 
\ref{RST-reducible-X:r}$\implies$\ref{RST-reducible-X:e}: For each $\mu\in L^X$, since $X\setminus X'$ is $\varphi$-reducible in RST, there exists $\mu'\in L^{X'}$ with
\[\ephi(\mu)=(\varphi_{X',Y})^\exists(\mu').\]
Thus
\begin{align*}
\ephi(\mu)&=(\varphi_{X',Y})^\exists(\mu')\\
&=(\varphi_{X',Y})^\exists(\varphi_{X',Y})^\forall(\varphi_{X',Y})^\exists(\mu')&((\varphi_{X',Y})^\exists\dv(\varphi_{X',Y})^\forall)\\
&=(\varphi_{X',Y})^\exists(\varphi_{X',Y})^\forall\ephi(\mu)\\
&=(\varphi_{X',Y})^\exists\left(\aphi\ephi(\mu)\right)_{X'}.&(\text{Lemma \ref{cd}\ref{restric}})
\end{align*}

\ref{RST-reducible-X:e}$\implies$\ref{RST-reducible-X:a}: Just note that
\[\ephi\aphi(\lam)=\ephi(\aphi(\lam))=(\varphi_{X',Y})^\exists\left(\aphi\ephi\aphi(\lam)\right)_{X'}=(\varphi_{X',Y})^\exists\left(\aphi(\lam)\right)_{X'}=(\varphi_{X',Y})^\exists(\varphi_{X',Y})^\forall(\lam)\]
for all $\lam\in L^Y$, where the last equality follows from Lemma \ref{cd}\ref{restric}.

\ref{RST-reducible-X:a}$\implies$\ref{RST-reducible-X:s}: By Lemma \ref{lem}\ref{lem-2}, it suffices to show that
\[\mu=\aphi\ephi(\underline{\mu_{X'}})\]
for all $\mu\in \CK\varphi$. Indeed, 
\begin{align*}
\mu&=\aphi\ephi\aphi\ephi(\mu)&(\mu\in\CK\varphi)\\
&=\aphi(\varphi_{X',Y})^\exists(\varphi_{X',Y})^\forall\ephi(\mu)\\
&=\aphi(\varphi_{X',Y})^\exists\left(\left((\aphi\ephi(\mu)\right)_{X'}\right)&(\text{Lemma \ref{cd}\ref{restric}})\\
&=\aphi(\varphi_{X',Y})^\exists(\mu_{X'})&(\mu\in\CK\varphi)\\
&=\aphi\ephi(\underline{\mu_{X'}}).&(\text{Lemma \ref{cd}\ref{underline}})
\end{align*}

\ref{RST-reducible-X:s}$\implies$\ref{RST-reducible-X:r}: For each $\mu\in L^X$, since $\CK(\tau,1_Y)$ is surjective and $\aphi\ephi(\mu)\in \CK\varphi$, there exists $\mu'\in \CK\varphi_{X',Y}\subseteq L^{X'}$ with 
\[\aphi\ephi(\mu)=\aphi\ephi(\underline\mu').\]
Then it follows from Lemma \ref{cd}\ref{underline} that
\[\ephi(\mu)=\ephi(\underline{\mu'})=(\varphi_{X',Y})^\exists(\mu'),\]
and consequently $X\setminus X'$ is $\varphi$-reducible in RST.
\end{proof}

\begin{prop} \label{RST-reducible-Y}
Let $(X,Y,\varphi)$ be an $L$-context and $Y'\subseteq Y$. The following statements are equivalent:
\begin{enumerate}[label={\rm(\roman*)}]
\item \label{RST-reducible-Y:r} $Y\setminus Y'$ is $\varphi$-reducible in RST.
\item \label{RST-reducible-Y:a} $\aphi\lambda=(\varphi_{X,Y'})^\forall\left(\ephi\aphi(\lambda)\right)_{Y'}$ for all $\lambda\in L^Y$.

\item \label{RST-reducible-Y:e} $\aphi\ephi=(\varphi_{X,Y'})^\forall(\varphi_{X,Y'})^\exists\colon L^X\to L^X$.
\item \label{RST-reducible-Y:i} $\CK(1_X,\nu)\colon\CK\varphi\to\CK\varphi_{X,Y'}$ is the identity function on $\CK\varphi=\CK\varphi_{X,Y'}$.
\end{enumerate}
\end{prop}

\begin{proof}
\ref{RST-reducible-Y:r}$\iff$\ref{RST-reducible-Y:a}$\iff$\ref{RST-reducible-Y:e} can be proved analogously to the equivalences of \ref{RST-reducible-X:r}, \ref{RST-reducible-X:e}, \ref{RST-reducible-X:a} in Proposition \ref{RST-reducible-X}.

\ref{RST-reducible-Y:e}$\implies$\ref{RST-reducible-Y:i}: $\CK\varphi=\CK\varphi_{X,Y'}$ is obvious and 
\[\CK(1_X,\nu)(\mu)=\aphi\ephi 1_X^{\rightarrow}(\mu)=\aphi\ephi(\mu)=\mu\]
for all $\mu\in \CK\varphi=\CK\varphi_{X,Y'}$.

\ref{RST-reducible-Y:i}$\implies$\ref{RST-reducible-Y:e}: By definition, $\CK\varphi$ and $\CK\varphi_{X,Y'}$ consist of fixed points of the maps 
\[\aphi \ephi\colon L^X\to L^X\quad\text{and}\quad(\varphi_{X,Y'})^\forall (\varphi_{X,Y'})^\exists\colon L^X\to L^X,\] 
respectively. Thus, if $\CK\varphi=\CK\varphi_{X,Y'}$, it follows from Proposition \ref{Fix}\ref{fix-ad} that, when restricting the codomain to the image, both $\aphi \ephi$ and $(\varphi_{X,Y'})^\forall (\varphi_{X,Y'})^\exists$ are the left adjoint of the same inclusion map, and therefore they must be equal.
\end{proof}

\begin{thm} \label{RST-reducible-reduct}
An $L$-context $(X',Y',\varphi_{X',Y'})$ is a reduct of $(X,Y,\varphi)$ in RST if, and only if, both $X\setminus X'$ and $Y\setminus Y'$ are $\varphi$-reducible in RST.
\end{thm}

\begin{proof}
``$\implies$'': Suppose that $(X',Y',\varphi_{X',Y'})$ is a reduct of $(X,Y,\varphi)$ in RST. Then both $F_1$ and $F_2$ are isomorphisms (see \eqref{F1-def} and \eqref{F2-def}). Thus, both the map
\[{\CK(\tau,1_Y)}\colon\varphi_{X',Y}\to \CK\varphi\]
and the inclusion map 
\[\CK\varphi_{X,Y'}\ \to/^(->/\CK\varphi\] 
are surjective; in particular, the inclusion map $\CK\varphi_{X,Y'}\to/^(->/\CK\varphi$ must be the identity map on $\CK\varphi_{X,Y'}=\CK\varphi$, and so is its left adjoint $\CK(1_X,\nu)$ (cf. Lemma \ref{lem}\ref{lem-1}). Therefore, $X\setminus X'$ is $\varphi$-reducible in RST by Proposition \ref{RST-reducible-X}, and $Y\setminus Y'$ is $\varphi$-reducible in RST by Proposition \ref{RST-reducible-Y}.

``$\impliedby$'': If both $X\setminus X'$ and $Y\setminus Y'$ are $\varphi$-reducible in RST, then for each $\lambda\in L^Y$, there exists $\lambda'\in L^{Y'}$ such that 
\[\aphi(\lambda)=(\varphi_{X,Y'})^\forall(\lambda').\] 
By Lemma \ref{cd}\ref{restric}, we have
\[(\varphi_{X',Y})^\forall(\lam)=\left(\aphi(\lambda)\right)_{X'}=\left((\varphi_{X,Y'})^\forall(\lambda')\right)_{X'}=(\varphi_{X',Y'})^\forall(\lambda').\]
Thus $Y\setminus Y'$ is $\varphi_{X',Y}$-reducible in RST. By Propositions \ref{RST-reducible-X} and \ref{RST-reducible-Y}, we deduce that
\[\CK(\tau,1_Y)\colon\CK\varphi_{X',Y}\to\CK\varphi\quad\text{and}\quad\CK(1_{X'},\nu)\colon\CK\varphi_{X',Y}\to\CK\varphi_{X',Y'}\]
are both isomorphisms. Thus, $(-)_{X'}\colon\CK\varphi\to\CK\varphi_{X',Y}$, the inverse of $\CK(\tau,1_Y)$, is an isomorphism (cf. Lemma \ref{lem}\ref{lem-2}). Therefore,
\[S_1=(\CK\varphi\to^{(-)_{X'}}\CK\varphi_{X',Y}\to^{\CK(1_X',J)}\CK\varphi_{X',Y'})\]
is an isomorphism, showing that $(X',Y',\varphi_{X',Y'})$ is a reduct of $(X,Y,\varphi)$ in RST.
\end{proof}

\section{Reducts of $L$-contexts: Formal concept analysis vs. rough set theory} \label{Reducts_FCA_RST}

It is well known that every complete $L$-lattice is isomorphic to the formal concept $L$-lattice of an $L$-context \cite{Bvelohlavek2004}. However, not every complete $L$-lattice is isomorphic to the property oriented concept $L$-lattice of an $L$-context:

\begin{thm} (See \cite[Theorem 5.3]{Lai2009}.)
Let $(L,*)$ be a complete residuated lattice. Then every complete $L$-lattice is isomorphic to the property oriented concept $L$-lattice of an $L$-context if, and only if, $L$ satisfies the law of double negation.
\end{thm}

In fact, in the case that $L$ satisfies the law of double negation, the formal concept $L$-lattice of an $L$-context $(X,Y,\varphi)$ is exactly the property oriented concept $L$-lattice of the $L$-context $(X,Y,\neg\varphi)$. Explicitly, since
\[\left((\neg\varphi)^\exists(\mu)\right)(y)=\bv\limits_{x\in X}\mu(x)*\neg\varphi(x,y)=\neg\Big(\bw\limits_{x\in X}\mu(x)\ra \varphi(x,y) \Big)=\neg\left(\varphi^\uparrow(\mu)(y)\right)\]
for all $\mu\in L^X$, $y\in Y$, it follows that
\begin{equation} \label{neg-phi-exists}
(\neg\varphi)^{\exists}=\neg\circ\uphi,\quad\text{and similarly,}\quad(\neg\varphi)^{\forall}=\dphi\circ\neg;
\end{equation}
consequently,
\begin{equation} \label{Mphi-Knegphi}
\Mphi=\Fix(\dphi\uphi)=\Fix((\neg\varphi)^{\forall}(\neg\varphi)^{\exists})=\CK(\neg\varphi).
\end{equation}

So, it is natural to ask whether reducts of $(X,Y,\varphi)$ in FCA correspond to reducts of $(X,Y,\neg\varphi)$ in RST. Our main result shows that the law of double negation also plays a crucial role in establishing the correspondence. Specifically, when $L$ satisfies the law of double negation, reducts of $L$-contexts in FCA are interdefinable with reducts of $L$-contexts in RST:

\begin{thm} \label{main}
Given a complete residuated lattice $L=(L,*)$, the following statements are equivalent:
\begin{enumerate}[label={\rm(\roman*)}]
\item \label{main:L} $L$ satisfies the law of double negation.
\item \label{main:iff} For each $L$-context $(X,Y,\varphi)$ and $X'\subseteq X$, $Y'\subseteq Y$, $(X',Y',\neg\varphi_{X',Y'})$ is a reduct of $(X,Y,\neg\varphi)$ in RST if, and only if, $(X',Y',\varphi_{X',Y'})$ is a reduct of $(X,Y,\varphi)$ in FCA.
\item \label{main:if} For each $L$-context $(X,Y,\varphi)$ and $X'\subseteq X$, $Y'\subseteq Y$, if $(X',Y',\neg\varphi_{X',Y'})$ is a reduct of $(X,Y,\neg\varphi)$ in RST, then $(X',Y',\varphi_{X',Y'})$ is a reduct of $(X,Y,\varphi)$ in FCA.
\end{enumerate}
\end{thm}

\begin{proof}
\ref{main:L}$\implies$\ref{main:iff}: We consider the map $E_1$ (see \eqref{E1-def}) for the $L$-context $(X,Y,\varphi)$ and the map $F_1$ (see \eqref{F1-def}) for the $L$-context $(X,Y,\neg\varphi)$. By \eqref{neg-phi-exists} and \eqref{Mphi-Knegphi}, we have
\[F_1(\mu')=(\neg\varphi)^\forall(\neg\varphi)^\exists(\underline{\mu'})=\dphi\uphi(\underline{\mu'})=E_1(\mu')\]
for all $\mu'\in \CM\varphi_{X',Y'}=\CK(\neg\varphi_{X',Y'})$. 
Hence, $F_1$ is an isomorphism if and only if so is $E_1$, and the conclusion thus follows.

\ref{main:iff}$\implies$\ref{main:if}: Trivial.

\ref{main:if}$\implies$\ref{main:L}: Let $a\in L$, and let $b\coloneqq a\ra 0$. Define an $L$-context $(\{x,y\},\{\star\},\varphi)$ with
\[\varphi(x,\star)=0\quad\text{and}\quad \varphi(y,\star)=a.\]
For any $\lambda\in L^{\{\star\}}$, note that
\begin{align*}
&(\neg\varphi)^\exists(\neg\varphi)^\forall(\lambda)\\
={}&(\neg\varphi(x,\star)*(\neg\varphi(x,\star)\ra\lambda(\star)))\vee(\neg\varphi(y,\star)*(\neg\varphi(y,\star)\ra\lambda(\star)))\\
={}&(1*(1\ra\lambda(\star)))\vee(b*(b\ra\lambda(\star)))\\
={}&\lambda(\star)&(\text{since}\ b*(b\ra\lambda(\star))\leq\lambda(\star)))\\
={}&1*(1\ra\lambda(\star))\\
={}&\neg\varphi(x,\star)*(\neg\varphi(x,\star)\ra\lambda(\star))\\
={}&{(\neg\varphi_{\{x\},\{\star\}})}^\exists{(\neg\varphi_{\{x\},\{\star\}})}^\forall(\lambda).
\end{align*}
Thus
\[(\neg\varphi)^\exists(\neg\varphi)^\forall={(\neg\varphi_{\{x\},\{\star\}})}^\exists{(\neg\varphi_{\{x\},\{\star\}})}^\forall\colon L^{\{\star\}} \to L^{\{\star\}},\] 
which necessarily forces $(\{x\},\{\star\},\neg\varphi_{\{x\},\{\star\}})$ to be a reduct of $(\{x,y\},\{\star\},\neg\varphi)$ in RST (see Proposition \ref{RST-reducible-X} and Theorem \ref{RST-reducible-reduct}). By hypothesis, $(\{x\},\{\star\},\varphi_{\{x\},\{\star\}})$ is a reduct of $(\{x,y\},\{\star\},\varphi)$ in FCA. Thus, by considering $\mu\in L^{\{\star\}}$ with $\mu(\star)=a$ we have
\begin{align*}
a&=((a\ra 0)\ra 0)\wedge a & (\text{since}\ a\leq (a\ra 0)\ra 0)\\
&=((a\ra 0)\ra 0)\wedge((a\ra a)\ra a)\\
&=((\mu(\star)\ra\varphi(x,\star))\ra\varphi(x,\star) )\wedge((\mu(\star)\ra\varphi(y,\star))\ra\varphi(y,\star))\\
&=\uphi\dphi(\mu)\\
&=({\varphi_{\{x\},\{\star\}}})^{\ua}({\varphi_{\{x\},\{\star\}}})^{\da}(\mu) & (\text{Proposition \ref{FCA-reducible-X} and Theorem \ref{FCA-reducible-reduct}})\\
&=(\mu(\star)\ra\varphi(x,\star))\ra\varphi(x,\star)\\
&=(a\ra 0)\ra 0\\
&=\neg\neg a.
\end{align*}
By the arbitrariness of $a$, we conclude that $L$ satisfies the law of double negation.
\end{proof}


\section{Examples} \label{Reducts_Examples}

In this section, we provide examples to illustrate the relationship between reducts of $L$-contexts in FCA and RST as discussed in the theorems above. 




\begin{exmp} \label{rational}
Let $\bbQ$ be the set of rational numbers. For $A,B\subseteq\bbQ$, the following statements are equivalent:
\begin{enumerate}[label=(\roman*)]
\item \label{rational:FCA} $(A,B,\leq)$ is a reduct of the (crisp) context $(\bbQ,\bbQ,\leq)$ in FCA;
\item \label{rational:RST} $(A,B,\not\leq)$ is a reduct of the (crisp) context $(\bbQ,\bbQ,\not\leq)$ in RST;
\item \label{rational:dense} both $A$ and $B$ are dense in $\bbQ$. 
\end{enumerate}
In fact, the equivalence of \ref{rational:FCA} and \ref{rational:RST} is an immediate consequence of Theorem \ref{main}. For the equivalence of \ref{rational:FCA} and \ref{rational:dense}, by Remark \ref{reducible} it is easy to see that
\[\bbQ\setminus A\ \text{is $\leq$-reducible in FCA}\iff\forall x\in\bbQ\setminus A\colon x=\bv A'\ \text{for some}\ A'\subseteq A;\]
that is, $A$ is dense in $\bbQ$. Similarly, $\bbQ\setminus B$ is $\leq$-reducible in FCA if and only if $B$ is dense in $\bbQ$. Hence, the conclusion follows from Theorem \ref{FCA-reducible-reduct}.
\end{exmp}

\begin{exmp} \label{topological}
Let $X$ be a topological space and $\CC$ the set of closed sets of $X$. 
For any subset $\CA\subseteq\CC$, the following statements are equivalent:
\begin{enumerate}[label=(\roman*)]
\item \label{topological:FCA} $(X,\CA,\in)$ is a reduct of the (crisp) context  $(X,\CC,\in)$ in FCA;
\item \label{topological:RST} $(X,\CA,\not\in)$ is a reduct of the (crisp) context  $(X,\CC,\not\in)$ in RST;
\item \label{topological:base} $\CA$ is a base for the closed sets of $X$; that is, for every $C\in\CC$, there exists $\CA'\subseteq\CA$ such that $C=\bigcap\CA'$.
\end{enumerate}
In fact, the equivalence of \ref{topological:FCA} and \ref{topological:RST} is an immediate consequence of Theorem \ref{main}. For the equivalence of \ref{topological:FCA} and \ref{topological:base}, note that
\[\in^{\da}(\CA')=\{x\in X\mid \forall A\in\CA'\colon x\in A\}=\bigcap\CA',\]
and consequently,
\begin{align*}
\CC\setminus\CA\ \text{is $\in$-reducible}&\iff\forall A\in \CC\setminus\CA\colon \in^{\da}(\{A\})=\in^{\da}(\CA')\ \text{for some}\ \CA'\subseteq\CA & (\text{Remark \ref{reducible}})\\
&\iff \forall A\in \CC\setminus\CA\colon A=\bigcap\CA'\ \text{for some}\ \CA'\subseteq\CA\\
&\iff \CA\ \text{is a base for the closed sets of}\ X.
\end{align*}
\end{exmp}

Theorem \ref{main} guarantees that, whenever $L$ does not satisfy the law of double negation, there exists an $L$-context $(X,Y, \varphi)$ such that $(X',Y', \neg \varphi_{X',Y'})$ is a reduct of $(X,Y, \neg \varphi)$ in RST, but $(X',Y', \varphi_{X',Y'})$ is not a reduct of $(X,Y, \varphi)$ in FCA. We provide several such examples below.

\begin{exmp} \label{exmp-three-chain}
Let $L=\{0<a<1\}$ be the three-chain. Then $L=(L,\wedge)$ is a frame that does not satisfy the law of double negation (see Example \ref{L-exmp}\ref{L-exmp:frame}), because
\[\neg\neg a=(a\ra 0)\ra 0=0\ra 0=1\neq a.\]
Define an $L$-context $(\{x,y\},\{\star\},\varphi)$ with
\[\varphi(x,\star)=0\quad\text{and}\quad \varphi(y,\star)=a.\]
It follows that 
\[\neg\varphi(x,\star)=0\ra 0=1\quad \text{and}\quad \neg\varphi(y,\star)=a\ra 0=0.\]
For any $\lam\in L^{\{\star\}}$, we have
\begin{align*}
(\neg\varphi)^\exists(\neg\varphi)^\forall(\lam)&=(\neg\varphi(x,\star)*(\neg\varphi(x,\star)\ra\lam(\star)))\vee (\neg\varphi(y,\star)*(\neg\varphi(y,\star)\ra\lam(\star)))\\
&=(1\wedge(1\ra\lam(\star)))\vee(0\wedge(0\ra\lam(\star)))\\
&=\lam(\star)\\
&=\neg\varphi(x,\star)*(\neg\varphi(x,\star)\ra\lam(\star))\\
&=(\neg\varphi_{\{x\},\{\star\}})^\exists(\neg\varphi_{\{x\},\{\star\}})^\forall(\lam).
\end{align*}
Then, we can see that $(\{x\},\{\star\},\neg\varphi_{\{x\},\{\star\}})$ is a reduct of $(\{x,y\},\{\star\},\neg\varphi)$ in RST, but $(\{x\},\{\star\},\varphi_{\{x\},\{\star\}})$ is not a reduct of $(\{x,y\},\{\star\},\varphi)$ in FCA, because by considering $\mu\in L^{\{\star\}}$ with $\mu(\star)=a$ we have
\[\uphi\dphi(\mu)=a\neq\neg\neg a=({\varphi_{\{x\},\{\star\}}})^{\ua}({\varphi_{\{x\},\{\star\}}})^{\da}(\mu).\]
\end{exmp}

\begin{exmp}
Let $[0,\infty]$ be the complete residuated lattice considered in Example \ref{L-exmp}\ref{L-exmp:Lawvere}. Define an $L$-context $(\{x,y\},\{\star\},\varphi)$ with
\[\varphi(x,\star)=\infty\quad\text{and}\quad\varphi(y,\star)=a\in(0,\infty).\]
By analogy with Example \ref{exmp-three-chain}, we obtain that $(\{x\},\{\star\},\neg\varphi_{\{x\},\{\star\}})$ is a reduct of $(\{x,y\},\{\star\},\neg\varphi)$ in RST. Let $\mu\in L^{\{\star\}}$. If $0<\mu(\star)\leq a$, then
\[\varphi^\ua\varphi^\da(\mu)(\star)=\big((\mu(\star)\ra\infty)\ra\infty \big)\vee\big((\mu(\star)\ra a)\ra a \big)=0\vee (a-(a-\mu(\star)))=\mu(\star)\]
and
\[\varphi_{\{x\},\{\star\}}^\ua\varphi_{\{x\},\{\star\}}^\da(\mu)(\star)=(\mu(\star)\ra\infty)\ra\infty=0.\]
Thus, $(\{x\},\{\star\},\varphi_{\{x\},\{\star\}})$ is not a reduct of $(\{x,y\},\{\star\},\varphi)$ in FCA.
\end{exmp}

\begin{exmp}
Let $[0,1]$ be equipped with the \emph{G{\"o}del t-norm} (also \emph{minimum t-norm}) $\wedge$. Then $([0,1],\wedge)$ does not satisfy the law of double negation (see Example \ref{L-exmp}\ref{L-exmp:BL}). Define an $L$-context $(\{x,y\},\{\star\},\varphi)$ with
\[\varphi(x,\star)=0\quad\text{and}\quad \varphi(y,\star)=a\in(0,1).\]
Similarly to Example \ref{exmp-three-chain}, $(\{x\},\{\star\},\neg\varphi_{\{x\},\{\star\}})$ is a reduct of $(\{x,y\},\{\star\},\neg\varphi)$ in RST. Let $\mu\in L^{\{\star\}}$. If $0<\mu(\star)\leq a$, then 
\[\varphi^\ua\varphi^\da(\mu)(\star)=\big((\mu(\star)\ra 0)\ra 0\big)\wedge\big((\mu(\star)\ra a)\ra a \big)=1\wedge (1\ra a)=a\]
and
\[\varphi_{\{x\},\{\star\}}^\ua\varphi_{\{x\},\{\star\}}^\da(\mu)(\star)=(\mu(\star)\ra 0)\ra 0=1.\]
Thus, $(\{x\},\{\star\},\varphi_{\{x\},\{\star\}})$ is not a reduct of $(\{x,y\},\{\star\},\varphi)$ in FCA.
\end{exmp}

\begin{exmp}
Let $\times$ be the \emph{product t-norm} on $[0,1]$. Then $([0,1],\times)$ does not satisfy the law of double negation (see Example \ref{L-exmp}\ref{L-exmp:BL}). Define an $L$-context $(\{x,y\},\{\star\},\varphi)$ with
\[\varphi(x,\star)=0\quad\text{and}\quad \varphi(y,\star)=a\in(0,1).\]
Analogously to Example \ref{exmp-three-chain}, $(\{x\},\{\star\},\neg\varphi_{\{y\},\{\star\}})$ is a reduct of $(\{x,y\},\{\star\},\neg\varphi)$ in RST. Let $\mu\in L^{\{\star\}}$. If $0<\mu(\star)\leq a$, then 
\[\varphi^\ua\varphi^\da(\mu)(\star)=\big((\mu(\star)\ra 0)\ra 0\big)\wedge\big((\mu(\star)\ra a)\ra a \big)=1\wedge (1\ra a)=a\]
and
\[\varphi_{\{x\},\{\star\}}^\ua\varphi_{\{x\},\{\star\}}^\da(\mu)(\star)=(\mu(\star)\ra 0)\ra 0=1.\]
Thus, $(\{x\},\{\star\},\varphi_{\{x\},\{\star\}})$ is not a reduct of $(\{x,y\},\{\star\},\varphi)$ in FCA.
\end{exmp}

\section{Algorithms} \label{Reducts_Algorithms}

Given an $L$-context $(X,Y,\varphi)$, we can programmatically determine whether an $L$-subcontext $(X',Y',\varphi_{X',Y'})$ is a reduct of $(X,Y,\varphi)$ in FCA or RST, as demonstrated in the following Algorithms \ref{algorithm_1} and \ref{algorithm_2}. The core idea is to verify the conditions specified by Theorems \ref{FCA-reducible-reduct} and \ref{RST-reducible-reduct} (also cf. Propositions \ref{FCA-reducible-X}, \ref{RST-reducible-X} and \ref{RST-reducible-Y}). 


\begin{algorithm}[!t] \label{algorithm_1}
    \caption{Determination of Reduct in FCA}
    \label{alg:fca_reduct}
    \BlankLine
    \KwIn{An $L$-context $(X, Y, \varphi)$; $L$-subcontext $(X', Y', \varphi_{X',Y'})$; $L$-lattice operations $(*, \ra, \wedge, \vee)$.}
    \KwOut{Boolean: \textbf{True} if $(X', Y', \varphi_{X',Y'})$ is an FCA reduct, \textbf{False} otherwise.}
    \BlankLine

    \tcp{Step 1: Check Y-Reducibility}
    $is\_Y\_reducible \leftarrow$ \textbf{True}\;
    \For{each $\mu \in L^X$}{
        \tcp{Compute original intent-extent closure $\dphi\uphi(\mu)$}
        $v_{orig} \leftarrow \dphi\uphi(\mu)$\;
        \tcp{Compute closure in subcontext with reduced property set $Y'$}
        $v_{sub} \leftarrow (\varphi_{X,Y'})^\da(\varphi_{X,Y'})^\ua(\mu)$\;
        
        \If{$v_{orig} \neq v_{sub}$}{
            $is\_Y\_reducible \leftarrow$ \textbf{False}\;
            \textbf{break}\;
        }
    }

    \BlankLine
    \tcp{Step 2: Check X-Reducibility}
    $is\_X\_reducible \leftarrow$ \textbf{True}\;
    \For{each $\lambda \in L^Y$}{
        \tcp{Compute original extent-intent closure $\uphi\dphi(\lambda)$}
        $w_{orig} \leftarrow \uphi\dphi(\lambda)$\;
        \tcp{Compute closure in subcontext with reduced object set $X'$}
        $w_{sub} \leftarrow (\varphi_{X',Y})^\ua(\varphi_{X',Y})^\da(\lambda)$\;

        \If{$w_{orig} \neq w_{sub}$}{
            $is\_X\_reducible \leftarrow$ \textbf{False}\;
            \textbf{break}\;
        }
    }

    \BlankLine
    \tcp{Final Verification}
    \Return $is\_Y\_reducible \land is\_X\_reducible$\;
\end{algorithm}

\begin{algorithm}[!t] \label{algorithm_2}
    \caption{Determination of Reduct in RST}
    \label{alg:rst_reduct_refined}
    \BlankLine
    \KwIn{An $L$-context $(X, Y, \varphi)$; $L$-subcontext $(X', Y', \varphi_{X',Y'})$;  $L$-lattice operations $(*, \ra, \wedge, \vee)$.}
    \KwOut{Boolean: \textbf{True} if $(X', Y', \varphi_{X',Y'})$ is an RST reduct, \textbf{False} otherwise.}
    \BlankLine

    \tcp{Step 1: Check Y-Reducibility}
    $is\_Y\_reducible \leftarrow$ \textbf{True}\;
    \For{each $\mu \in L^X$ }{
        $v_{orig} \leftarrow \aphi\ephi(\mu)$\;
        $v_{sub} \leftarrow (\varphi_{X,Y'})^\forall(\varphi_{X,Y'})^\exists(\mu)$\;
        
        \If{$v_{orig} \neq v_{sub}$}{
            $is\_Y\_reducible \leftarrow$ \textbf{False}\;
            \textbf{break}\;
        }
    }

    \BlankLine
    \tcp{Step 2: Check X-Reducibility}
    $is\_X\_reducible \leftarrow$ \textbf{True}\;
    \For{each  $\lambda \in L^Y$}{
        $w_{orig} \leftarrow \ephi\aphi(\lambda)$\;
        $w_{sub} \leftarrow (\varphi_{X',Y})^\exists(\varphi_{X',Y})^\forall(\lambda)$\;

        \If{$w_{orig} \neq w_{sub}$}{
            $is\_X\_reducible \leftarrow$ \textbf{False}\;
            \textbf{break}\;
        }
    }

    \BlankLine
    \tcp{Final Verification}
    \Return $is\_Y\_reducible \land is\_X\_reducible$\;
\end{algorithm}

The time complexity of these algorithms depend on the size of the concept lattices involved. Suppose that $X$, $Y$ and $L$ are both finite. Let $|X|$ be the number of objects, $|Y|$ the number of properties, and $|L|$ the cardinality of the complete residuated lattice.

To check the reducibility of $Y \setminus Y'$, the algorithm iterates through all $L$-subsets $\mu \in L^X$. The number of such sets is $|L|^{|X|}$. Inside the loop, the closure operators $\dphi\uphi$ and $\aphi\ephi$ involve matrix-vector multiplications over the residuated lattice, taking $O(|X| \cdot |Y|)$ time. Therefore, the worst-case time complexity is $O(|L|^{|X|} \cdot |X| \cdot |Y| + |L|^{|Y|} \cdot |X| \cdot |Y|)$.

It is important to note that these algorithms are designed for verification purposes --- to rigorously check if a given subcontext qualifies as a reduct according to Definitions \ref{reduct-FCA} and \ref{reduct-RST-def} --- rather than for heuristic data mining of large datasets. Consequently, they are most suitable for theoretical validation or datasets where $|X|$ and $|Y|$ are small, or where $L$ is a small finite chain.

\begin{prop}[Correctness of Algorithm \ref{algorithm_1}] 
Algorithm \ref{algorithm_1} returns {\bf True} if and only if $(X', Y', \varphi_{X',Y'})$ is a reduct of $(X, Y, \varphi)$ in FCA.
\end{prop}

\begin{proof}
 Step 1 of Algorithm \ref{algorithm_1} verifies the condition $\dphi\uphi = (\varphi_{X,Y'})^{\da}(\varphi_{X,Y'})^{\ua}$, which corresponds exactly to the necessary and sufficient condition for the $\varphi$-reducibility of $Y \setminus Y'$ established in Proposition \ref{FCA-reducible-X}\ref{FCA-reducible-X:Y}. Similarly, 
 Step 2 verifies the condition for the $\varphi$-reducibility of $X \setminus X'$ established in \ref{FCA-reducible-X}\ref{FCA-reducible-X:X}. Final verification returns the conjunction of these results. By Theorem \ref{FCA-reducible-reduct}, the intersection of these two reducibilities is equivalent to the definition of a reduct in FCA.
\end{proof}

Similarly, we have the following parallel proposition for Algorithm \ref{algorithm_2}:

\begin{prop}[Correctness of Algorithm \ref{algorithm_2}] 
Algorithm \ref{algorithm_2} returns {\bf True} if and only if $(X', Y', \varphi_{X',Y'})$ is a reduct of $(X, Y, \varphi)$ in RST.
\end{prop}

In the remainder of this section, we present worked examples to illustrate the proposed algorithms:

\begin{exmp}
\label{ex:algorithm_trace}
Let $L=\{0,0.5,1\}$ be the complete residuated lattice equipped with the Lukasiewicz structure, i.e., 
\[a * b = 0\vee(a+b-1)\quad\text{and}\quad a \ra b =1\wedge(1-a+b).\] 
Let $X=\{x_1, x_2\}$ and $Y=\{y_1, y_2\}$. Consider an $L$-context $(X, Y, \varphi)$ defined by the matrix
\[
\varphi = 
\begin{pmatrix}
1 & 0.5 \\
0.5 & 1
\end{pmatrix}.
\]
We apply Algorithm \ref{algorithm_1} to verify whether the subcontext $(X, \{y_1\}, \varphi_{X, \{y_1\}})$ (obtained by removing the property $y_2$) is a reduct in FCA. This requires checking if $Y \setminus \{y_1\} = \{y_2\}$ is $\varphi$-reducible. According to the algorithm (and Proposition \ref{FCA-reducible-X}), we must verify if 
\begin{equation} \label{phi-mu-x-y1}
\varphi^{\downarrow}\varphi^{\uparrow}(\mu) = (\varphi_{X, \{y_1\}})^{\downarrow}(\varphi_{X, \{y_1\}})^{\uparrow}(\mu)
\end{equation}
holds for all $\mu \in L^X$.

Let us trace the iteration for the specific $L$-subset 
\[\mu\colon X\to L,\quad\mu(x_1)=1,\quad\mu(x_2)=0\]
(represented as the vector $(1, 0)$):

\begin{enumerate}[label=(\arabic*)]
\item Compute the LHS of \eqref{phi-mu-x-y1} (original context):
First, compute $\varphi^{\uparrow}(\mu)$:
\begin{align*}
\varphi^{\uparrow}(\mu)&=\mu\ra\varphi=(1,0)\ra\begin{pmatrix}
1 & 0.5 \\
0.5 & 1
\end{pmatrix}\\
&=\big((1 \ra 1) \wedge (0 \ra 0.5),(1 \ra 0.5) \wedge (0 \ra 1)\big)=(1 \wedge 1,0.5 \wedge 1)=(1,0.5).
\end{align*}
Next, compute $\varphi^{\downarrow}(1, 0.5)$:
\begin{align*}
\varphi^{\downarrow}(1,0.5)&=(1,0.5)\ra\varphi^T=(1,0.5)\ra\begin{pmatrix}
1 & 0.5 \\
0.5 & 1
\end{pmatrix}\\
&=\big((1 \ra 1) \wedge (0.5 \ra 0.5),(1 \ra 0.5) \wedge (0.5 \ra 1)\big)=(1 \wedge 1,0.5 \wedge 1)=(1,0.5).
\end{align*}
    LHS result: $\varphi^{\downarrow}\varphi^{\uparrow}(\mu)=\varphi^{\downarrow}(1, 0.5)= (1, 0.5)$.

    \item Compute the RHS of \eqref{phi-mu-x-y1} (reduced context): Here, we only consider the column $y_1$. The restriction is
    \[\varphi_{X, \{y_1\}} =\begin{pmatrix}
1 \\
0.5
\end{pmatrix}.\]
First, compute $(\varphi_{X, \{y_1\}})^{\uparrow}(\mu)$:
\[(\varphi_{X, \{y_1\}})^{\uparrow}(\mu)(y_1)=\mu\ra\varphi_{X, \{y_1\}}=(1,0)\ra\begin{pmatrix}
1 \\
0.5
\end{pmatrix}= (1 \ra 1) \wedge (0 \ra 0.5) = 1\wedge 1=1.\]
Next, compute $(\varphi_{X, \{y_1\}})^{\downarrow}(1)$:
\[(\varphi_{X, \{y_1\}})^{\downarrow}(1)=(1)\ra\varphi_{X, \{y_1\}}^T=(1)\ra(1,0.5)=(1,0.5).\]
    RHS result: $(\varphi_{X, \{y_1\}})^{\downarrow}(\varphi_{X, \{y_1\}})^{\uparrow}(\mu) =(\varphi_{X, \{y_1\}})^{\downarrow}(1) = (1, 0.5)$.
\end{enumerate}

So, the equation \eqref{phi-mu-x-y1} holds for this specific $\mu$. The algorithm proceeds to check the remaining $3^{|X|} - 1 = 8$ $L$-subsets of $X$. If the equality holds for all, and the corresponding condition for objects (checking $\lambda \in L^Y$) also holds, the algorithm returns \textbf{True}.
\end{exmp}

\begin{exmp} \label{wine}
    We apply the proposed algorithms to a real-world dataset to examine the behavior of reducts in FCA and RST under different fuzzy logic structures. The experiments are conducted on the Wine dataset from the UCI Machine Learning Repository. We randomly selected a subset consisting of 5 representative objects and 7 chemical attributes. The specific samples and their fuzzy attribute values are presented in Table \ref{tab:raw_wine_data}. The set of attributes is 
    \[Y = \{\text{Alcohol, Malic Acid, Ash, Alcalinity of ash, Magnesium, Phenols, Flavanoids}\}.\]

\begin{table}[htbp]
\centering
\caption{Original Attribute Values of Selected Samples from UCI Wine Dataset}
\label{tab:raw_wine_data}
\begin{tabular}{lccccccc}
\toprule
Object ID & Alcohol & Malic Acid & Ash & Alcalinity of ash & Magnesium & Phenols & Flavanoids \\
\midrule
$x_1$ (Index 17)  & 13.83 & 1.57 & 2.62 & 20.0 & 115 & 2.95 & 3.40 \\
$x_2$ (Index 61)  & 12.64 & 1.36 & 2.02 & 16.8 & 100 & 2.02 & 1.41 \\
$x_3$ (Index 68)  & 13.34 & 0.94 & 2.36 & 17.0 & 110 & 2.53 & 1.30 \\
$x_4$ (Index 132) & 12.81 & 2.31 & 2.40 & 24.0 & 98  & 1.15 & 1.09 \\
$x_5$ (Index 153) & 13.23 & 3.30 & 2.28 & 18.5 & 98  & 1.80 & 0.83 \\
\bottomrule
\end{tabular}
\end{table}

The raw values were first normalized to the unit interval $[0, 1]$ and then quantized into a finite chain 
\[L = \{0, 0.25, 0.5, 0.75, 1\}\]
to facilitate the computation. We conducted the tests across two kinds of $L$-contexts, where $L$ is equipped with the minimum and the {\L}ukasiewicz structure, respectively. We verify the reducibility of each attribute $y_i \in Y$, i.e., check if $Y \backslash \{y_i\}$ is reducible.

\begin{table}[htbp]
\centering
\caption{Comparative Reducibility Analysis: G\"{o}del vs. {\L}ukasiewicz Logic}
\label{tab:attribute_removal_test}
\begin{tabular}{lcccccc}
\toprule
\multirow{2}{*}{\textbf{Attribute Under Test}} & \multicolumn{3}{c}{\textbf{G\"{o}del}} & \multicolumn{3}{c}{\textbf{{\L}ukasiewicz}} \\
\cmidrule(lr){2-4} \cmidrule(lr){5-7}
& FCA & RST & NRST & FCA & RST & NRST \\
\midrule
Alcohol           & NO  & NO  & \textbf{YES} & \textbf{YES} & NO  & \textbf{YES} \\
Malic acid        & NO  & NO  & NO           & NO  & NO  & NO           \\
Ash               & NO  & \textbf{YES} & \textbf{YES} & NO  & \textbf{YES} & NO           \\
Alcalinity of ash & NO  & NO  & \textbf{YES} & NO  & NO  & NO           \\
Magnesium         & \textbf{YES} & NO  & \textbf{YES} & NO  & NO  & NO           \\
Total phenols     & NO  & NO  & NO           & NO  & NO  & NO           \\
Flavanoids        & NO  & NO  & NO           & NO  & NO  & NO           \\
\midrule
\textbf{Total Reducible} & 1 & 1 & 4 & 1 & 1 & 1 \\
\bottomrule
\multicolumn{7}{l}{\footnotesize * NRST denotes NEG\_RST (reduction in RST under the negation operator).}
\end{tabular}
\end{table}

The experimental results in Table \ref{tab:attribute_removal_test} demonstrate how the choice of the specific residuated lattice determines the relationship between reduction methods:
\begin{itemize}
    \item Divergence under G\"{o}del logic: Under the G\"{o}del structure, the three reduction criteria show no formal inclusion or equivalence. For instance, while Ash is reducible in NRST, it is not reducible in FCA.  Magnesium is reducible in FCA but not reducible in RST. 
    \item Consistency under {\L}ukasiewicz logic: Under the {\L}ukasiewicz structure, FCA and NEG\_RST yield identical results for all tested attributes.  This empirically validates our established Theorem \ref{main}, which proves that when the law of double negation is satisfied, the reduction criteria of FCA and NEG\_RST are equivalent.
\end{itemize}
 
\end{exmp}

\section{Conclusions and future work}

The main contributions of this paper are threefold:
\begin{itemize}
\item We provide a simplified characterization of reducts in $L$-contexts based on FCA using infomorphisms. 
\item We introduce a corresponding definition for reducts in $L$-contexts based on RST using comparison maps between property-oriented concept lattices.
\item Most significantly, we prove in Theorem \ref{main} that these two notions of reduction --- one stemming from FCA and the other from RST --- are interdefinable via negation if and only if the underlying lattice $L$ satisfies the law of double negation.
\end{itemize}
These findings provide a solid theoretical foundation for reduction in fuzzy data analysis, clarifying when methods from one field can be validly applied to the other. For future work, several directions are promising:
\begin{itemize}
\item Algorithmic optimization: The algorithms presented in Section \ref{Reducts_Algorithms} are verification-based and have exponential complexity. Developing heuristic algorithms to find optimal minimal reducts in polynomial time for specific subclasses of residuated lattices is a practical necessity.
\item Three-way decisions: Investigating how these reduction methods interact with three-way decision theories in fuzzy environments.
\item Heterogeneous data: Extending these results to heterogeneous contexts where objects and properties may take values in different lattice structures.
\end{itemize}

\section*{Data availability}

The source code implementing the proposed Algorithms \ref{algorithm_1} and \ref{algorithm_2}, as well as the experiments presented in Example \ref{wine}, is available at [\url{https://github.com/ChanYuxu/Fuzzy-reduction-FCA-and-RST}]. The Wine dataset used in this example is available in the UCI Machine Learning Repository [\url{http://archive.ics.uci.edu/ml/datasets/Wine}].

\section*{Acknowledgement}

We acknowledge the support of National Natural Science Foundation of China (No. 12071319) and the Fundamental Research Funds for the Central Universities (No. 2021SCUNL202). We are grateful for helpful remarks received from the anonymous referee which help us improve the presentation of this paper significantly.





\begin{thebibliography}{10}

\bibitem{Alsina2006}
C.~Alsina, M.~J. Frank, and B.~Schweizer.
\newblock {\em Associative Functions: Triangular Norms and Copulas}.
\newblock World Scientific, Singapore, 2006.

\bibitem{Barr1991}
M.~Barr.
\newblock $*$-{Autonomous} categories and linear logic.
\newblock {\em Mathematical Structures in Computer Science}, 1:159--178, 1991.

\bibitem{Bvelohlavek2002}
R.~B{\v e}lohl{\' a}vek.
\newblock {\em Fuzzy Relational Systems: Foundations and Principles}, volume~20
  of {\em IFSR International Series on Systems Science and Engineering}.
\newblock Kluwer Academic Publishers, Dordrecht, 2002.

\bibitem{Bvelohlavek2004}
R.~B{\v e}lohl{\' a}vek.
\newblock Concept lattices and order in fuzzy logic.
\newblock {\em Annals of Pure and Applied Logic}, 128(1-3):277--298, 2004.

\bibitem{BenitezCaballero2020}
M.~J. Ben{\'\i}tez-Caballero, J.~Medina, E.~Ram{\'\i}rez-Poussa, and
  D.~{\'S}l{\k e}zak.
\newblock Rough-set-driven approach for attribute reduction in fuzzy formal
  concept analysis.
\newblock {\em Fuzzy Sets and Systems}, 391:117--138, 2020.

\bibitem{Chang1958}
C.~C. Chang.
\newblock Algebraic analysis of many valued logics.
\newblock {\em Transactions of the American Mathematical Society},
  88(2):467--490, 1958.

\bibitem{Davey2002}
B.~A. Davey and H.~A. Priestley.
\newblock {\em Introduction to Lattices and Order}.
\newblock Cambridge University Press, Cambridge, second edition, 2002.

\bibitem{Duntsch2002}
I.~D{\"u}ntsch and G.~Gediga.
\newblock Modal-style operators in qualitative data analysis.
\newblock In V.~Kumar, S.~Tsurnoto, X.~Wu, P.~S. Yu, and N.~Zhong, editors,
  {\em Proceedings of the 2002 IEEE International Conference on Data Mining},
  pages 155--162. IEEE Computer Society, Los Alamitos, 2002.

\bibitem{Ganter2007}
B.~Ganter.
\newblock Relational {Galois} connections.
\newblock In S.~O. Kuznetsov and S.~Schmidt, editors, {\em Formal Concept
  Analysis}, volume 4390 of {\em Lecture Notes in Computer Science}, pages
  1--17. Springer, Berlin--Heidelberg, 2007.

\bibitem{Ganter1999}
B.~Ganter and R.~Wille.
\newblock {\em Formal Concept Analysis: Mathematical Foundations}.
\newblock Springer, Berlin--Heidelberg, 1999.

\bibitem{Hajek1998}
P.~H{\'a}jek.
\newblock {\em Metamathematics of Fuzzy Logic}, volume~4 of {\em Trends in
  Logic}.
\newblock Springer, Dordrecht, 1998.

\bibitem{Klement2000}
E.~P. Klement, R.~Mesiar, and E.~Pap.
\newblock {\em Triangular Norms}, volume~8 of {\em Trends in Logic}.
\newblock Springer, Dordrecht, 2000.

\bibitem{Krotzsch2005}
M.~Kr{\"o}tzsch, P.~Hitzler, and G.~Zhang.
\newblock Morphisms in context.
\newblock In F.~Dau, M.-L. Mugnier, and G.~Stumme, editors, {\em Conceptual
  Structures: Common Semantics for Sharing Knowledge}, volume 3596 of {\em
  Lecture Notes in Computer Science}, pages 223--237. Springer,
  Berlin--Heidelberg, 2005.

\bibitem{Lai2009}
H.~Lai and D.~Zhang.
\newblock Concept lattices of fuzzy contexts: Formal concept analysis vs. rough
  set theory.
\newblock {\em International Journal of Approximate Reasoning}, 50(5):695--707,
  2009.

\bibitem{Mi2004}
J.-S. Mi, W.-Z. Wu, and W.-X. Zhang.
\newblock Approaches to knowledge reduction based on variable precision rough
  set model.
\newblock {\em Information Sciences}, 159(3):255--272, 2004.

\bibitem{Pawlak1982}
Z.~Pawlak.
\newblock Rough sets.
\newblock {\em International Journal of Computer $\&$ Information Sciences},
  11(5):341--356, 1982.

\bibitem{Poelmans2014}
J.~Poelmans, D.~I. Ignatov, S.~O. Kuznetsov, and G.~Dedene.
\newblock Fuzzy and rough formal concept analysis: a survey.
\newblock {\em International Journal of General Systems}, 43(2):105--134, 2014.

\bibitem{Polkowski2002}
L.~Polkowski.
\newblock {\em Rough Sets: Mathematical Foundations}, volume~15 of {\em
  Advances in Intelligent and Soft Computing}.
\newblock Physica-Verlag, Heidelberg, 2002.

\bibitem{Pratt1995}
V.~Pratt.
\newblock Chu spaces and their interpretation as concurrent objects.
\newblock In J.~Leeuwen, editor, {\em Computer Science Today}, volume 1000 of
  {\em Lecture Notes in Computer Science}, pages 392--405. Springer,
  Berlin--Heidelberg, 1995.

\bibitem{Rosenthal1990}
K.~I. Rosenthal.
\newblock {\em Quantales and their Applications}, volume 234 of {\em Pitman
  research notes in mathematics series}.
\newblock Longman, Harlow, 1990.

\bibitem{Shen2014}
L.~Shen.
\newblock {\em Adjunctions in Quantaloid-enriched Categories}.
\newblock PhD thesis, Sichuan University, Chengdu, 2014.

\bibitem{Shen2016a}
L.~Shen, Y.~Tao, and D.~Zhang.
\newblock Chu connections and back diagonals between
  $\mathcal{Q}$-distributors.
\newblock {\em Journal of Pure and Applied Algebra}, 220(5):1858--1901, 2016.

\bibitem{Shen2013a}
L.~Shen and D.~Zhang.
\newblock Categories enriched over a quantaloid: {Isbell} adjunctions and {Kan}
  adjunctions.
\newblock {\em Theory and Applications of Categories}, 28(20):577--615, 2013.

\bibitem{Shen2013}
L.~Shen and D.~Zhang.
\newblock The concept lattice functors.
\newblock {\em International Journal of Approximate Reasoning}, 54(1):166--183,
  2013.

\bibitem{Stubbe2005}
I.~Stubbe.
\newblock Categorical structures enriched in a quantaloid: categories,
  distributors and functors.
\newblock {\em Theory and Applications of Categories}, 14(1):1--45, 2005.

\bibitem{Stubbe2014}
I.~Stubbe.
\newblock An introduction to quantaloid-enriched categories.
\newblock {\em Fuzzy Sets and Systems}, 256:95--116, 2014.

\bibitem{Thangavel2009}
K.~Thangavel and A.~Pethalakshmi.
\newblock Dimensionality reduction based on rough set theory: A review.
\newblock {\em Applied Soft Computing}, 9(1):1--12, 2009.

\bibitem{Wang2008}
X.~Wang and W.~Zhang.
\newblock Relations of attribute reduction between object and property oriented
  concept lattices.
\newblock {\em Knowledge-Based Systems}, 21(5):398--403, 2008.

\bibitem{Wasilewski2023}
P.~Wasilewski, J.~Kacprzyk, and S.~Zadrożny.
\newblock Reduction of binary attributes: Rough set theory versus formal
  concept analysis.
\newblock In A.~Campagner, O.~Urs~Lenz, S.~Xia, D.~Ślęzak, J.~Wąs, and
  J.~Yao, editors, {\em Rough Sets}, pages 46--61, Cham, 2023. Springer.

\bibitem{Wei2010}
L.~Wei and J.-J. Qi.
\newblock Relation between concept lattice reduction and rough set reduction.
\newblock {\em Knowledge-Based Systems}, 23(8):934--938, 2010.

\bibitem{Wu2009}
W.-Z. Wu, Y.~Leung, and J.-S. Mi.
\newblock Granular computing and knowledge reduction in formal contexts.
\newblock {\em IEEE Transactions on Knowledge and Data Engineering},
  21(10):1461--1474, 2009.

\bibitem{Yao2004}
Y.~Yao.
\newblock Concept lattices in rough set theory.
\newblock In {\em Proceedings of 2004 Annual Meeting of the North American
  Fuzzy Information Processing Society (NAFIPS 2004)}, volume~2, pages
  796--801. IEEE, 2004.

\bibitem{Zhang2025}
Q.~Zhang, J.~Qi, L.~Wei, and S.~Zhao.
\newblock Attribute combination reduction in formal concept analysis: A
  theoretical characterization.
\newblock {\em International Journal of Approximate Reasoning}, 186:109498,
  2025.

\bibitem{Zhang2005a}
W.~Zhang, L.~Wei, and J.~Qi.
\newblock Attribute reduction theory and approach to concept lattice.
\newblock {\em Science in China Series F: Information Sciences},
  48(6):713--726, 2005.

\bibitem{Zhang2003a}
W.-X. Zhang, J.-S. Mi, and W.-Z. Wu.
\newblock Approaches to knowledge reductions in inconsistent systems.
\newblock {\em International Journal of Intelligent Systems}, 18(9):989--1000,
  2003.

\bibitem{Zhao2023a}
S.~Zhao, J.~Qi, J.~Li, and L.~Wei.
\newblock Concept reduction in formal concept analysis based on representative
  concept matrix.
\newblock {\em International Journal of Machine Learning and Cybernetics},
  14(4):1147--1160, 2023.

\end{thebibliography}

\end{document}